\documentclass[journal]{IEEEtran}

\usepackage{cite}          
\usepackage{amsmath,amssymb,amsfonts} 
\usepackage{graphicx}      
\usepackage{textcomp}      
\usepackage{xcolor}        

\usepackage{algorithm}     
\usepackage{algorithmic}   

\usepackage{array}
\usepackage{booktabs}      
\usepackage{multirow}      

\usepackage[caption=false,font=footnotesize,labelfont=sf,textfont=sf]{subfig}

\usepackage{url}           
\usepackage{verbatim}      
\usepackage{stfloats}      
\usepackage{balance}       
\usepackage{tablefootnote}
\usepackage[hidelinks]{hyperref}

\newif\ifmarked
\markedfalse   

\newcommand{\rev}[1]{%
  \ifmarked
    \textcolor{blue}{#1}%
  \else
    #1%
  \fi
}

\begin{document}

\title{Rotated Robustness: A Training-Free Defense against Bit-Flip Attacks on Large Language Models}

\author{
    Deng~Liu,~\IEEEmembership{Student~Member,~IEEE},
    and~Song~Chen,~\IEEEmembership{Member,~IEEE}
    \thanks{This work was supported in part by the National Natural Science Foundation of China under Grant Nos. 62672470 and 92473114, and in part by the Strategic Priority Research Program of the CAS under Grant Nos. XDB0660101, XDB0660100, and XDB0660000. (Corresponding author: Song Chen.)}
    \thanks{Deng Liu and Song Chen are with the School of Integrated Circuits, University of Science and Technology of China, Hefei 230026, China (e-mail: ld153@mail.ustc.edu.cn; songch@ustc.edu.cn).}
}

\maketitle

\begin{abstract}
Bit-flip corruption of quantized weights poses a serious reliability threat to Large Language Models (LLMs), as even a small number of weight faults can trigger catastrophic model degradation. We show that such failures can be strongly amplified when corrupted weights are coupled to high-sensitivity activation channels. \rev{Based on this observation, we propose Rotated Robustness (RoR), a training-free sensitivity-aware hierarchical rotation that applies matched orthogonal transformations to activations and weights, redistributing sensitive channel contributions across feature dimensions while preserving the underlying linear mapping in exact arithmetic.}

\rev{Across eight evaluated LLMs, RoR matches or exceeds all compared methods in the number of cumulative PBS flips sustained before PPL exceeds 100 within the 100-flip evaluation horizon. Under BER-based random bit-flip injection, no RoR failures are observed in the evaluated 1,000-trial experiments for the tested models. Under defense-aware SPoF reconstruction, RoR increases the cost of reproducing specified one-bit catastrophic failures to thousands of deployed INT8 bit changes. RoR also preserves downstream-task utility under PBS perturbations. On NVIDIA H200 at batch size 8, RoR incurs 0.097\%--0.584\% persistent storage overhead, 5.2\%--14.1\% Prefill overhead, and 0.5\%--9.8\% Decode overhead across four representative models. These results show that sensitivity-aware hierarchical rotation provides a practical approach to improving the bit-flip robustness of quantized LLM Linear weights.}
\end{abstract}

\section{Introduction}
\label{sec:intro}

Large Language Models (LLMs) have rapidly expanded from high-performance GPU clusters~\cite{narayanan2021efficient} to increasingly diverse deployment platforms~\cite{liu2024mobilellm}, supporting applications such as autonomous agents~\cite{park2023generative} and financial analysis~\cite{wu2023bloomberggpt}. Representative model families, including Llama~2/3~\cite{touvron2023llama,dubey2024llama}, Qwen~\cite{bai2023qwen}, and GPT-4~\cite{achiam2023gpt}, demonstrate strong reasoning capability at billion-parameter scale~\cite{wei2022chain}. As these models are deployed on physical hardware, however, their parameters become exposed to memory-level reliability and security threats.

Rowhammer~\cite{kim2014flipping} can induce bit flips in DRAM by repeatedly activating aggressor rows, while end-to-end studies such as BitMine~\cite{zhang2021bitmine} demonstrate that software-induced DRAM faults can be triggered at selected memory locations by reverse-engineering address mappings. These capabilities enable Bit-Flip Attacks (BFAs)~\cite{rakin2019bit,hong2019terminal} that directly corrupt stored model parameters. Unlike input-level adversarial examples, BFAs alter the model itself and can cause severe functional degradation with only a few carefully selected flips. ONEFLIP~\cite{li2025rowhammer} shows that a single bit flip can inject a stealthy backdoor, while AttentionBreaker~\cite{das2024attentionbreaker} reports catastrophic degradation of an INT8-quantized LLaMA3-8B model after only three weight-bit flips. \rev{More recently, Flip-Agent~\cite{wang2026flipagent} extends targeted BFAs to LLM-based agents, showing that bit-level parameter corruption can manipulate not only final outputs but also tool invocations.} These results highlight the growing practical risk of parameter-level faults in LLM deployments. \rev{In this work, we focus on corruptions of eligible INT8 weight bits in Transformer Linear layers.}

Defending billion-parameter LLMs against BFAs is challenging because existing defenses often introduce additional inference cost or require model modification. Detection-based methods~\cite{li2021radar,liu2023neuropots,guo2021modelshield,liu2024alberta,ahmed2024nn} add runtime integrity checking or error detection, whose deployment cost depends strongly on the execution substrate and memory-access pattern. Weight-robustness methods instead modify the model or its representation: robust retraining and fine-tuning~\cite{zhou2024sar,li2020defending} require additional optimization at LLM scale and may affect clean-model utility, while encoding-based schemes~\cite{velvcicky2024deepncode,liu2022generating} introduce extra decoding or data-transformation operations on the inference path. This motivates a training-free defense that directly reduces the sensitivity of deployed weight-bit faults while retaining practical inference efficiency.

\noindent\textit{Our Insight: Outlier Alignment and Amplification.}
We identify the interaction between corrupted weights and extreme activation channels as a key amplification mechanism behind catastrophic bit-flip failures in LLMs. Transformer activations frequently contain persistent outlier channels whose magnitudes can substantially exceed those of ordinary channels~\cite{dettmers2022gpt3,xiao2023smoothquant}. When a corrupted weight directly couples to such a high-sensitivity channel, the resulting error can be strongly amplified and propagated through subsequent layers, producing the observed Single Point of Failure (SPoF) behavior.

\noindent\textit{Our Approach: RoR.}
\rev{We propose \textbf{Rotated Robustness (RoR)}, a training-free sensitivity-aware hierarchical rotation. RoR uses offline activation sensitivity to guide how strongly different feature channels are mixed, redistributing the influence of high-sensitivity channels across feature dimensions and reducing their susceptibility to localized weight-bit corruption. The rotation is an exact-arithmetic reparameterization before quantization; in practical INT8 deployment, the transformed weights are re-quantized and the resulting finite-precision impact is evaluated empirically. Orthogonal rotations have also been explored in LLM quantization, e.g., QuaRot~\cite{ashkboos2024quarot}, to suppress activation outliers and improve low-bit quantization rather than to defend against bit-flip attacks. RoR instead develops a Householder-motivated, sensitivity-aware hierarchical rotation specifically for robustness against localized weight-bit corruption.\footnote{The RoR source code is publicly available at \url{https://github.com/liudeng1/RoR-Defense}.}}

Our contributions are threefold:
\begin{itemize}
    \item \rev{\textit{Mechanism and Defense.}
    We identify outlier--weight coupling as a key amplification mechanism in LLM bit-flip failures and introduce RoR, a training-free sensitivity-aware hierarchical rotation that reduces the concentration of high-sensitivity channel influence without retraining.}

    \item \rev{\textit{Robustness and Generalization.}
    Across eight evaluated LLMs, RoR achieves the best or tied-best PPL$>100$ threshold-crossing budget under gradient-guided Progressive Bit Search. It also suppresses catastrophic degradation under random bit-flip attacks, preserves downstream-task utility under replayed PBS perturbations, and increases the reconstruction cost of specified one-bit SPoFs to thousands of deployed INT8 bit changes.}

    \item \rev{\textit{Measured Deployment Overhead.}
    On NVIDIA H200 at batch size 8, RoR incurs 0.097\%--0.584\% persistent-storage overhead, 5.2\%--14.1\% Prefill overhead, and 0.5\%--9.8\% one-token Decode overhead across four representative models, providing a practical robustness--efficiency trade-off for W8A16 LLM deployment.}
\end{itemize}

\section{Background and Threat Model}
\label{sec:background_threat}

\subsection{LLM Inference and Quantization}
\begin{figure}[t]
    \centering
    \includegraphics[width=0.95\linewidth]{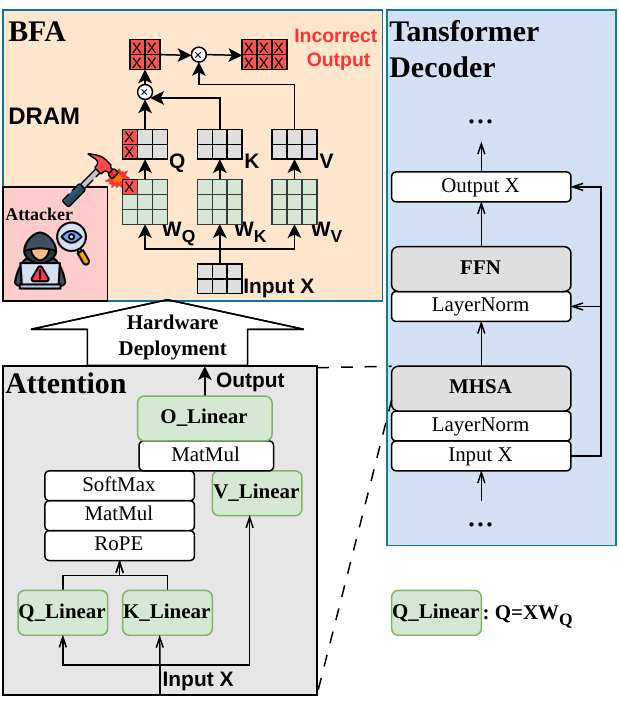}
    \caption{\rev{\textbf{Bit-flip error propagation in Transformers.} A physical fault or targeted attack corrupts one or more bits in a quantized Linear weight stored in memory. The resulting localized error can propagate and be amplified through subsequent Transformer computations, potentially causing catastrophic output degradation.}}
    \label{fig:transformer_attacker}
\end{figure}

Modern Large Language Models (LLMs) are fundamentally built upon the Transformer architecture~\cite{vaswani2017attention}. A standard Transformer block comprises Multi-Head Self-Attention (MHSA) and a Feed-Forward Network (FFN), whose computation is dominated by dense linear transformations. Given an activation matrix $\mathbf{X}\in\mathbb{R}^{T\times d}$, a linear layer computes
\begin{equation}
    \mathbf{Y}=\mathbf{X}\mathbf{W},
\end{equation}
where $\mathbf{W}\in\mathbb{R}^{d\times d'}$ is the weight matrix. As illustrated in Fig.~\ref{fig:transformer_attacker}, MHSA projects activations into Queries, Keys, and Values through learnable linear weights, while the FFN similarly relies on large linear transformations.

To reduce inference memory and computation cost, Post-Training Quantization (PTQ) is widely used in LLM deployment~\cite{frantar2022gptq}. 
{\ifmarked\color{blue}\fi
In the deployment setting evaluated in this work, Transformer Linear weights are symmetrically quantized to INT8 using per-output-channel FP32 scales. For an element $w_{ij}$ associated with output channel $j$,
\begin{equation}
    q_{ij}
    =
    \operatorname{Clamp}
    \left(
        \left\lfloor \frac{w_{ij}}{s_j} \right\rceil,
        -128,127
    \right),
\end{equation}
where $s_j$ is the corresponding FP32 scale and $q_{ij}$ is stored in INT8 two's-complement representation. The deployed value is reconstructed as $\hat{w}_{ij}=s_jq_{ij}$.
}

\subsection{Threat Model: Hardware Faults and Adversary Capabilities}
\label{subsec:threat_model}

Bit-level corruption of quantized model parameters can directly alter LLM inference. Two physical mechanisms motivate the threat model considered here. First, Rowhammer~\cite{kim2014flipping} can induce targeted DRAM bit flips by repeatedly activating aggressor rows, and end-to-end studies such as BitMine~\cite{zhang2021bitmine} show that vulnerable locations can be exploited after reverse-engineering memory address mappings. Second, stochastic soft errors can arise from transient physical disturbances such as cosmic radiation~\cite{baumann2005radiation} or aggressive undervolting~\cite{chang2017understanding}. Based on attacker knowledge and attack guidance, we consider three hierarchical adversary settings.

\rev{Throughout all three settings, the writable fault domain is fixed to the INT8 two's-complement payload of Transformer Linear-layer weights. The three settings differ only in the attacker's knowledge and attack guidance rather than in which model storage locations are writable. Whenever a bit-flip count or attack cost is reported, it is measured as the XOR Hamming distance between the clean and corrupted deployed INT8 weight representations. For cumulative attacks such as PBS, this equals the number of distinct persistent INT8 bit changes.}

\rev{Bit-flip vulnerability can also arise in other parameter classes. TrojBits studies inference-time bit-flip attacks on word embeddings and discusses embedding-specific mitigation strategies, including rare-token embedding replacement and embedding-layout randomization~\cite{alghanim2023trojbits}. Quantization scale factors constitute another distinct attack surface: corrupting a single scale can simultaneously affect a group of de-quantized weights, and recent work shows that existing targeted-BFA and critical-bit protection remains insufficient against such attacks~\cite{Wang_2025_CVPR}. Normalization parameters such as LayerNorm weights are likewise structurally distinct from dense Linear weights and are not included in the Linear-weight fault domain studied here; this limitation is discussed later. Accordingly, this work studies direct corruption of quantized Transformer Linear weights as a distinct and practically important fault domain, rather than claiming full-model protection against arbitrary memory corruption.}

\rev{Auxiliary defense metadata, if present, is outside the writable fault domain evaluated in this work. We do not claim robustness to direct corruption of such metadata; the present study is restricted to the Transformer Linear INT8 weight-bit domain defined above.}

\textit{1) Black-Box (Stochastic):}
The adversary has no knowledge of the model architecture or parameters. Corruptions are unguided and stochastic, represented by BER-based random bit-flip injection over the defined weight-bit domain. This setting also captures hardware soft errors such as Single Event Upsets (SEUs) from cosmic radiation~\cite{baumann2005radiation} or instability induced by aggressive voltage scaling~\cite{chang2017understanding}.

\rev{\textit{2) Gray-Box (Targeted):}
The attacker knows the model architecture and has access to the deployed quantized weight values and gradients with respect to those deployed weights, but not the internal defense configuration. Thus, for transformed defenses such as RoR, the attacker directly targets the transformed and re-quantized INT8 weight representation. This enables gradient-guided or heuristic targeted attacks, such as Progressive Bit Search (PBS)~\cite{rakin2019bit} and AttentionBreaker~\cite{das2024attentionbreaker}, to iteratively identify vulnerable weight bits within the same writable fault domain.}

\rev{\textit{3) White-Box (Defense-Aware):}
The attacker additionally knows the complete defense algorithm and its instantiated configuration, together with the model weights and gradients. This allows the attacker to exploit defense information when constructing targeted attacks, while the writable fault domain remains the same quantized Linear-layer weight bits defined above. In particular, we consider defense-aware attempts to reproduce Single Point of Failure (SPoF) behavior, where a highly localized weight perturbation can trigger catastrophic model degradation~\cite{guo2025sbfa}. This setting therefore evaluates targeted weight-bit corruption under full defense knowledge without extending the writable domain to other model parameters or auxiliary metadata. We analyze the mechanism underlying SPoF behavior in Sec.~\ref{sec:motivation}.}

\subsection{Defense Objectives}
\label{subsec:defense_objectives}

Practical Bit-Flip Attacks are constrained by the physical characteristics of the underlying fault mechanism. Rowhammer-class exploits, for example, can induce only a sparse and spatially constrained set of flips in vulnerable memory locations. Motivated by these constraints and the deployment requirements of LLMs, an effective defense should satisfy four objectives:

\textit{1) Reliability (Black-Box):}
The defense should mitigate catastrophic failures under stochastic, unguided random bit-flip errors.

\rev{\textit{2) Attack Cost (Gray/White-Box):}
The defense should reduce the effectiveness of sparse targeted bit flips and increase the attack effort required to induce severe degradation or reproduce highly destructive failure modes.}

\rev{\textit{3) Utility Preservation:}
The defense should preserve the model's reasoning and generation quality, with its clean finite-precision impact explicitly measured.}

\textit{4) System Efficiency:}
The defense should limit both inference-latency and persistent-storage overhead in practical deployment.

\section{Motivation}
\label{sec:motivation}

\subsection{Rare Catastrophic Failures under Random Bit Flips}

Most prior BFA studies emphasize targeted, gradient-guided attacks~\cite{rakin2019bit,yao2020deephammer}, while stochastic random bit flips are often expected to be less damaging because large models contain substantial parameter redundancy. Our fault-injection results show that this redundancy does not eliminate rare catastrophic failures.

\begin{figure}[t]
    \centering
    \includegraphics[width=0.9\linewidth]{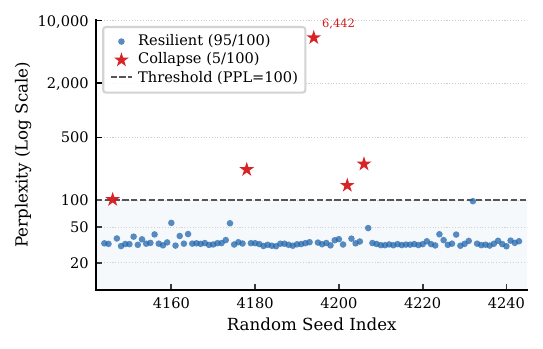}
    \caption{\rev{\textbf{Rare catastrophic failures under BER-based random bit-flip injection.} The figure shows a 100-seed window from the OPT-125M scan at BER $3{\times}10^{-4}$. Most trials retain near-baseline PPL, whereas five trials exceed the PPL$=100$ threshold; the most severe case reaches PPL 6,442.}}
    \label{fig:spof_phenomenon}
\end{figure}

\rev{We perform a BER-based random bit-flip scan on \texttt{OPT-125M} with $\mathrm{BER}=3{\times}10^{-4}$ over the eligible INT8 Transformer Linear weight-bit domain. Across 500 random seeds, 26 trials exceed the operational catastrophic-failure threshold of PPL$>100$ (5.2\%), while most trials remain close to the normal PPL range. The most severe trial reaches a PPL of 6,442. Fig.~\ref{fig:spof_phenomenon} visualizes a representative 100-seed window (seeds 4145--4244), in which 5 of 100 trials exceed the same threshold. These observations show that random bit-flip attacks can occasionally produce severe model degradation despite the large number of parameters.}

\rev{The BER trial itself may contain many sampled bit flips. To determine whether the observed collapse can be attributed to a highly localized vulnerability, we further localize the realized fault set in representative catastrophic trials and replay the isolated faults individually. This process identifies single-bit faults that can reproduce catastrophic degradation; we refer to such verified one-bit vulnerabilities as \emph{Single Points of Failure (SPoFs)}. The existence of such SPoFs motivates a defense that reduces sensitivity to localized weight corruption rather than relying solely on parameter redundancy.}

\subsection{Alignment of SPoFs with Activation Outliers}
\label{subsec:spatial_alignment}

\begin{figure}[t]
    \centering
    \includegraphics[width=0.75\linewidth]{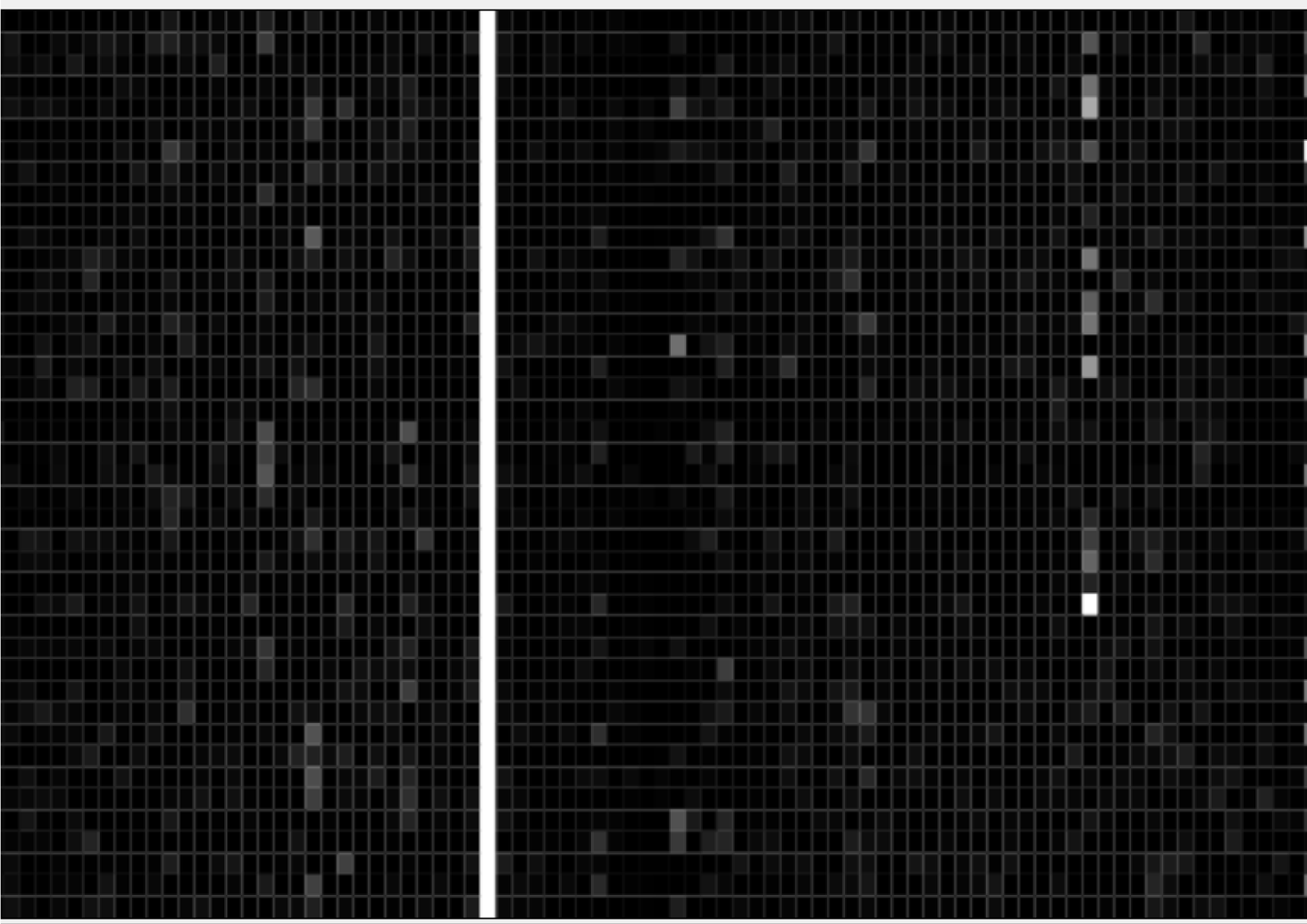}
    \caption{\textbf{SPoF alignment with an activation outlier.} Activations in \texttt{OPT-125M} \texttt{layer2.fc1} show that channel 706 reaches substantially larger magnitudes than surrounding channels. Faults coupled to this channel can therefore induce disproportionately large output perturbations.}
    \label{fig:fatal_alignment}
\end{figure}

To understand why an isolated bit can cause such severe degradation, we recursively localize the sampled fault set in representative catastrophic trials and examine the corresponding activation statistics. The localized faults affect weights coupled to the same input activation channel, channel 706, in \texttt{OPT-125M}. As visualized in Fig.~\ref{fig:fatal_alignment}, this channel is an extreme activation outlier: its magnitude reaches approximately 6, whereas surrounding channels are typically within roughly $[-0.2,0.2]$.

For a linear layer $\mathbf{Y}=\mathbf{X}\mathbf{W}$, a perturbation $\Delta W_{j,k}$ induces
\begin{equation}
    \Delta y_{i,k}=x_{i,j}\Delta W_{j,k},
\label{eq:perturbation}
\end{equation}
where $x_{i,j}$ is the activation of channel $j$ for token $i$. Thus, for tokens where $|x_{i,j}|$ is unusually large, the same weight perturbation produces a proportionally larger output error. A high-order INT8 bit flip can therefore become especially damaging when it affects a weight coupled to an extreme activation channel.

In the traced cases, the verified SPoFs are coupled to the same extreme activation channel. This empirical alignment provides the central motivation for reducing the concentration of high-sensitivity channel influence rather than treating all feature channels uniformly.

\section{Methodology}
\label{sec:method}

{\ifmarked\color{blue}\fi
In this section, we present Rotated Robustness (RoR), a training-free sensitivity-aware hierarchical rotation that disperses the influence of high-sensitivity activation channels across feature dimensions. RoR profiles channel sensitivity from clean calibration data, assigns deeper mixing to more sensitive channels, motivates this redistribution through Householder geometry, and realizes it efficiently through a hierarchy-consistent butterfly structure.
}

\subsection{Activation-Channel Sensitivity Characterization}
\label{subsec:vulnerability}

Consider a linear layer $\mathbf{Y}=\mathbf{X}\mathbf{W}$, where $\mathbf{X}\in\mathbb{R}^{N\times d_{in}}$ and $\mathbf{W}\in\mathbb{R}^{d_{in}\times d_{out}}$. We refer to the $j$-th column $\mathbf{X}_{:,j}$ as \emph{activation channel $j$}; under this matrix convention, it is coupled to the corresponding weight row $\mathbf{W}_{j,:}$. As established in Sec.~\ref{sec:motivation} (Eq.~\ref{eq:perturbation}), a perturbation $\Delta W_{j,k}$ produces
\begin{equation}
\label{eq:perturbation_method}
    \Delta y_{i,k}=x_{i,j}\Delta W_{j,k}.
\end{equation}
{\ifmarked\color{blue}\fi
For a fixed weight perturbation, the largest immediate disturbance over the activation sequence is therefore
\begin{equation}
\label{eq:local_amplification}
    \max_i |\Delta y_{i,k}|
    =|\Delta W_{j,k}|\,\|\mathbf{X}_{:,j}\|_{\infty}.
\end{equation}
Equation~\eqref{eq:local_amplification} shows that the peak magnitude of activation channel $j$ acts as a channel-dependent amplification factor for a perturbation in the corresponding weight row. This is a local linear-layer relation rather than a global upper bound on end-to-end model error.
}

Accordingly, we define the \emph{channel sensitivity score}
\begin{equation}
\label{eq:sensitivity_score}
    m_j=\|\mathbf{X}_{:,j}\|_{\infty}.
\end{equation}
The activation outliers observed in Sec.~\ref{subsec:spatial_alignment} correspond to the high-sensitivity end of this continuous channel-wise spectrum. Let $\mu$ and $\sigma$ denote the mean and standard deviation of $\{m_j\}_{j=1}^{d_{in}}$. We use the strict sensitivity threshold
\begin{equation}
\label{eq:threshold}
    \tau_{\alpha}
    =\max\!\left(\mu+\alpha\sigma,\;2\mu,\;1.0\right),
\end{equation}
and fix $\alpha=6$ throughout our evaluation.
{\ifmarked\color{blue}\fi
To extend strict outlier identification to a graded sensitivity spectrum, we further define
\begin{equation}
\label{eq:beta_threshold}
    \tau_{\beta}=\beta\tau_{\alpha},
    \qquad 0<\beta<1.
\end{equation}
Channels with $m_j>\tau_{\alpha}$ are termed high-sensitivity channels, those with $\tau_{\beta}<m_j\le\tau_{\alpha}$ intermediate-sensitivity channels, and the remainder low-sensitivity channels. Fig.~\ref{fig:ror_sensitivity} illustrates the sensitivity ranking and thresholding process.
}

\subsection{Sensitivity-Aware Rotated Robustness}
\label{subsec:framework}

\noindent\textbf{Orthogonal reparameterization.}
Let $\mathbf{Q}\in\mathbb{R}^{d_{in}\times d_{in}}$ satisfy $\mathbf{Q}^{\top}\mathbf{Q}=\mathbf{I}$. Then
\begin{equation}
\label{eq:orthogonal_equivalence}
\begin{aligned}
    \mathbf{Y}
    &=\mathbf{X}\mathbf{W}
    =\mathbf{X}(\mathbf{Q}\mathbf{Q}^{\top})\mathbf{W}\\
    &=(\mathbf{X}\mathbf{Q})(\mathbf{Q}^{\top}\mathbf{W})
    =\mathbf{X}_{\mathrm{RoR}}\mathbf{W}_{\mathrm{RoR}},
\end{aligned}
\end{equation}
where $\mathbf{X}_{\mathrm{RoR}}=\mathbf{X}\mathbf{Q}$ and $\mathbf{W}_{\mathrm{RoR}}=\mathbf{Q}^{\top}\mathbf{W}$.
\rev{Thus, a matched orthogonal transform redistributes activation features while preserving the underlying linear mapping in exact arithmetic before quantization. In INT8 deployment, the transformed weights are re-quantized after offline fusion, and the resulting clean-performance impact is evaluated empirically in Sec.~\ref{sec:evaluation}. The remaining question is how to construct $\mathbf{Q}$ so that high-sensitivity channel contributions are dispersed without imposing the same mixing workload on every channel.}

\noindent\textbf{Householder dispersion.}
A Householder reflection provides a direct geometric construction for redistributing one concentrated feature direction. Let $\mathbf{e}_j$ be the $j$-th canonical basis vector and define the equal-magnitude direction
\begin{equation}
\label{eq:householder_target}
    \mathbf{u}=\frac{1}{\sqrt{d_{in}}}\mathbf{1},
    \qquad
    \mathbf{v}_j=
    \frac{\mathbf{e}_j-\mathbf{u}}
         {\|\mathbf{e}_j-\mathbf{u}\|_2}.
\end{equation}
The Householder reflector~\cite{householder1958unitary}
\begin{equation}
\label{eq:householder_original}
    \mathbf{H}_j
    =\mathbf{I}-2\mathbf{v}_j\mathbf{v}_j^{\top}
\end{equation}
is orthogonal and symmetric, and satisfies
\begin{equation}
\label{eq:householder_mapping}
    \mathbf{e}_j^{\top}\mathbf{H}_j=\mathbf{u}^{\top}.
\end{equation}
Therefore, for an isolated channel contribution $x_j\mathbf{e}_j^{\top}$, a full-dimensional Householder reflection spreads its magnitude uniformly over all $d_{in}$ feature dimensions, reducing the magnitude carried by each feature dimension from $|x_j|$ to $|x_j|/\sqrt{d_{in}}$.

{\ifmarked\color{blue}\fi
The same Householder principle can be applied locally. In a two-dimensional subspace,
\begin{equation}
\label{eq:local_householder}
    \mathbf{H}_2
    =\frac{1}{\sqrt{2}}
    \begin{bmatrix}
        1 & 1\\
        1 & -1
    \end{bmatrix},
    \qquad
    \mathbf{H}_2^{\top}\mathbf{H}_2=\mathbf{I}_2,
\end{equation}
and
\begin{equation}
\label{eq:pairwise_operation}
    (a,b)\mathbf{H}_2
    =\left(
        \frac{a+b}{\sqrt{2}},
        \frac{a-b}{\sqrt{2}}
    \right).
\end{equation}
In particular, $(a,0)$ becomes $(a/\sqrt{2},a/\sqrt{2})$. We use this 2-D Householder operation as the local butterfly mixing primitive in the hierarchical construction below, thereby connecting the original Householder dispersion mechanism to an efficient structured implementation.
}

{\ifmarked\color{blue}\fi
\noindent\textbf{Sensitivity ranking and requested mixing depth.}
For clarity, we first assume $d_{in}=2^L$. Let $\pi=(\pi_1,\ldots,\pi_{d_{in}})$ denote the descending sensitivity ranking,
\begin{equation}
\label{eq:sensitivity_ranking}
    m_{\pi_1}\ge m_{\pi_2}\ge\cdots\ge m_{\pi_{d_{in}}},
\end{equation}
and define the corresponding permutation matrix
\begin{equation}
\label{eq:permutation_matrix}
    \mathbf{P}
    =\left[\mathbf{e}_{\pi_1},\mathbf{e}_{\pi_2},\ldots,
    \mathbf{e}_{\pi_{d_{in}}}\right].
\end{equation}
Hence, $\mathbf{X}\mathbf{P}$ places activation channels in descending sensitivity order.

Each original activation channel $j$ is assigned a requested mixing depth according to its sensitivity score. Let
\begin{equation}
\label{eq:kmin}
    k_{\min}=\max\!\left\{1,\left\lfloor\frac{L}{2}\right\rfloor-1\right\}
\end{equation}
denote the minimum structural mixing depth. We define
\begin{equation}
\label{eq:adaptive_depth}
r_j=
\begin{cases}
    L,
    & m_j>\tau_{\alpha},\\[1mm]
    k_{\min},
    & m_j\le\tau_{\beta},\\[1mm]
    k_{\min}
    +\left\lceil
    (L-k_{\min})
    \dfrac{\log(m_j/\tau_{\beta})}
          {\log(\tau_{\alpha}/\tau_{\beta})}
    \right\rceil,
    & \tau_{\beta}<m_j\le\tau_{\alpha}.
\end{cases}
\end{equation}
Thus, high-sensitivity channels request full-depth mixing, low-sensitivity channels retain the structural minimum, and intermediate-sensitivity channels receive progressively deeper mixing as $m_j$ increases.

The requested mixing depths cannot in general be realized independently because each local butterfly operation acts on a complete hierarchical group. We therefore construct the smallest \emph{hierarchy-consistent} plan satisfying all requests. Let
\begin{equation}
\label{eq:realized_depth}
    d_j\ge r_j
\end{equation}
denote the realized mixing depth of original channel $j$. The ranked realized-depth sequence is $[d_{\pi_1},\ldots,d_{\pi_{d_{in}}}]$. The requested mixing depth $r_j$ is a planning quantity, whereas the realized mixing depth $d_j$ determines the actual transform and its deployment representation.
}

\begin{figure}[!t]
    \centering
    \subfloat[Channel-sensitivity characterization.]{%
        \IfFileExists{figures/Fig4a.pdf}{%
            \includegraphics[width=0.47\linewidth]{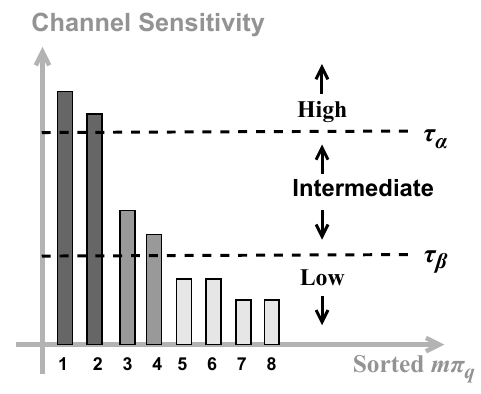}%
        }{%
            \fbox{\parbox[c][1.30in][c]{0.43\linewidth}{\centering
            Placeholder for Fig.~4(a)\\
            $m_{\pi_q}$, $\tau_{\alpha}$, $\tau_{\beta}$}}
        }%
        \label{fig:ror_sensitivity}%
    }%
    \hfill
    \subfloat[Hierarchy-consistent mixing plan.]{%
        \IfFileExists{figures/Fig4b.pdf}{%
            \includegraphics[width=0.47\linewidth]{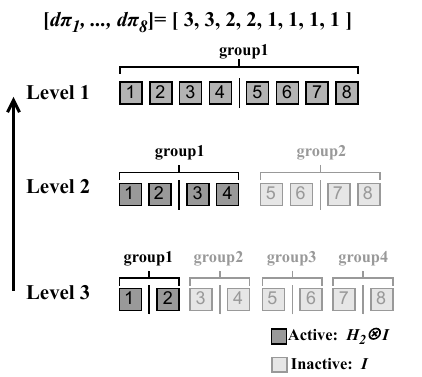}%
        }{%
            \fbox{\parbox[c][1.30in][c]{0.43\linewidth}{\centering
            Placeholder for Fig.~4(b)\\
            active/inactive hierarchy levels}}
        }%
        \label{fig:ror_hierarchy}%
    }
    \caption{\rev{\textbf{Sensitivity-aware hierarchical construction of RoR.}
    (a) Activation channels are ranked by the sensitivity score and partitioned into high-, intermediate-, and low-sensitivity regions using $\tau_{\alpha}$ and $\tau_{\beta}$.
    (b) The realized mixing depths determine which hierarchical groups are active. The illustrated eight-channel example uses $[d_{\pi_1},\ldots,d_{\pi_8}]=[3,3,2,2,1,1,1,1]$; active groups apply local Householder-based butterfly mixing, whereas inactive groups apply the identity.}}
    \label{fig:adaptive_rotation}
    \label{fig:householder_vis}
\end{figure}

{\ifmarked\color{blue}\fi
\noindent\textbf{Hierarchy-consistent Householder transform.}
We index hierarchy levels by $\ell=1,\ldots,L$, where $\ell=1$ is the coarsest root level and $\ell=L$ is the finest pairwise level. As illustrated in Fig.~\ref{fig:ror_hierarchy}, the sensitivity-ranked feature space is organized as a binary hierarchy: the root contains all $2^L$ ranked features, and each successive level bisects every parent group into two contiguous child groups. Consequently, level $\ell$ contains $2^{\ell-1}$ hierarchical groups, each with $2^{L-\ell+1}$ ranked features.

Let $\mathcal{G}_{\ell,g}$ denote the ranked feature positions belonging to group $g$ at level $\ell$:
\begin{equation}
\label{eq:hierarchical_group}
\mathcal{G}_{\ell,g}
=
\left\{
(g-1)2^{L-\ell+1}+1,\ldots,
 g2^{L-\ell+1}
\right\},
\end{equation}
where $g=1,\ldots,2^{\ell-1}$. During planning, a group must be active at level $\ell$ whenever at least one channel in that group requests mixing depth $\ell$ or deeper. After the minimal hierarchy-consistent construction, this condition is equivalently expressed as
\begin{equation}
\label{eq:group_activation_equiv}
    \max_{q\in\mathcal{G}_{\ell,g}} r_{\pi_q}\ge\ell
    \quad\Longleftrightarrow\quad
    \min_{q\in\mathcal{G}_{\ell,g}} d_{\pi_q}\ge\ell.
\end{equation}
Thus, each hierarchical group is either active as a whole or inactive as a whole at a given level.

The transform of group $g$ at level $\ell$ is defined from the realized mixing depths as
\begin{equation}
\label{eq:level_block}
\mathbf{B}_{\ell,g}=
\begin{cases}
    \mathbf{H}_2\otimes\mathbf{I}_{2^{L-\ell}},
    & \displaystyle
      \min_{q\in\mathcal{G}_{\ell,g}} d_{\pi_q}\ge\ell,\\[2mm]
    \mathbf{I}_{2^{L-\ell+1}},
    & \text{otherwise}.
\end{cases}
\end{equation}
An active group therefore performs $2^{L-\ell}$ independent local 2-D Householder operations between corresponding features in its two halves, forming one level of the hierarchical butterfly transform; an inactive group applies the identity. The complete transform at level $\ell$ is
\begin{equation}
\label{eq:level_matrix}
    \mathbf{B}_{\ell}
    =\operatorname{blkdiag}
    \left(
        \mathbf{B}_{\ell,1},\ldots,
        \mathbf{B}_{\ell,2^{\ell-1}}
    \right).
\end{equation}
Every $\mathbf{B}_{\ell}$ is orthogonal. The online implementation executes the finest active level first and proceeds toward the root; under the row-vector convention of Eq.~\eqref{eq:orthogonal_equivalence}, the complete hierarchical butterfly transform is
\begin{equation}
\label{eq:b_levels}
    \mathbf{B}=\mathbf{B}_{L}\mathbf{B}_{L-1}\cdots\mathbf{B}_{1}.
\end{equation}
Once the ranked realized-depth vector $[d_{\pi_1},\ldots,d_{\pi_{d_{in}}}]$ is known, the active/inactive state of every hierarchical group is uniquely determined. For example, Fig.~\ref{fig:ror_hierarchy} uses $[3,3,2,2,1,1,1,1]$: the root group is active at Level~1, only the first four-feature group is active at Level~2, and only the first pair is active at Level~3. The execution order is Level~3 $\rightarrow$ Level~2 $\rightarrow$ Level~1.

The implementation additionally applies a fixed seeded diagonal sign transform
\begin{equation}
\label{eq:sign_matrix}
    \mathbf{D}=\operatorname{diag}(s_1,\ldots,s_{d_{in}}),
    \qquad s_j\in\{-1,+1\}.
\end{equation}
Here $\mathbf{D}$ is defined in the sensitivity-ranked feature space. The complete layer-wise RoR transform is
\begin{equation}
\label{eq:q_factorization}
    \mathbf{Q}=\mathbf{P}\mathbf{B}\mathbf{D}.
\end{equation}
Because $\mathbf{P}$, every $\mathbf{B}_{\ell}$, and $\mathbf{D}$ are orthogonal,
\begin{equation}
\label{eq:q_orthogonality}
    \mathbf{Q}^{\top}\mathbf{Q}=\mathbf{I}.
\end{equation}
The sign transform changes only feature signs and therefore does not alter the magnitude-based dispersion described below. For non-power-of-two widths, the ranked channels are deterministically partitioned into contiguous power-of-two segments, and the same construction is applied independently within each segment; the resulting block-diagonal composition remains orthogonal.
}

{\ifmarked\color{blue}\fi
\noindent\textbf{Proposition 1 (Channel-wise dispersion).}
Consider a nonzero contribution $x_j\mathbf{e}_j^{\top}$ from original activation channel $j$. If its realized mixing depth is $d_j$, then it participates in the nested active hierarchy levels $\ell=1,\ldots,d_j$. Because these levels are executed from fine to coarse, each active level expands the contribution into the companion half of the next larger group. Consequently,
\begin{equation}
\label{eq:channel_dispersion}
    \left|\operatorname{supp}
    \!\left(x_j\mathbf{e}_j^{\top}\mathbf{Q}\right)\right|
    =2^{d_j},
    \qquad
    \left\|x_j\mathbf{e}_j^{\top}\mathbf{Q}\right\|_{\infty}
    =|x_j|\,2^{-d_j/2}.
\end{equation}
Moreover, orthogonality preserves its $\ell_2$ energy,
\begin{equation}
\label{eq:channel_energy}
    \left\|x_j\mathbf{e}_j^{\top}\mathbf{Q}\right\|_2
    =|x_j|.
\end{equation}
The first relation follows by induction from Eq.~\eqref{eq:pairwise_operation}: an isolated channel contribution initially occupies one feature dimension; each additional active hierarchy level doubles the number of affected feature dimensions and multiplies every resulting magnitude by $1/\sqrt{2}$. The factor $\mathbf{P}$ only reorders features, while $\mathbf{D}$ only changes signs, so neither changes the support size or absolute magnitude.

Combining Eq.~\eqref{eq:channel_dispersion} with the sensitivity score in Eq.~\eqref{eq:sensitivity_score} yields
\begin{equation}
\label{eq:sensitivity_dispersion}
    \max_i
    \left\|
        x_{i,j}\mathbf{e}_j^{\top}\mathbf{Q}
    \right\|_{\infty}
    =m_j\,2^{-d_j/2}.
\end{equation}
This relation gives the mathematical meaning of sensitivity-aware mixing-depth assignment: increasing $d_j$ reduces the maximum contribution from channel $j$ carried by any single feature dimension by the exact factor $2^{-d_j/2}$. When $d_j=L$, the contribution has equal magnitude $|x_j|/\sqrt{d_{in}}$ over all feature dimensions, matching the equal-magnitude dispersion of the full-dimensional Householder construction, although the sign pattern need not be identical.

Importantly, Eq.~\eqref{eq:sensitivity_dispersion} is a \emph{channel-wise contribution} guarantee. For a general dense activation vector, transformed contributions from different original channels may add constructively within the same feature dimension. We therefore do not claim a universal inequality of the form $\|\mathbf{x}\mathbf{Q}\|_{\infty}\le\|\mathbf{x}\|_{\infty}$ for every $\mathbf{x}$. Instead, RoR directly reduces the concentration of high-sensitivity channel contributions across feature dimensions identified by Eq.~\eqref{eq:local_amplification}.
}

\subsection{Deployment and Overhead Analysis}
\label{subsec:deployment}
\label{subsec:deployment_overhead}

\begin{figure}[!t]
    \centering
    \IfFileExists{figures/Fig5.pdf}{%
        \includegraphics[width=0.78\linewidth]{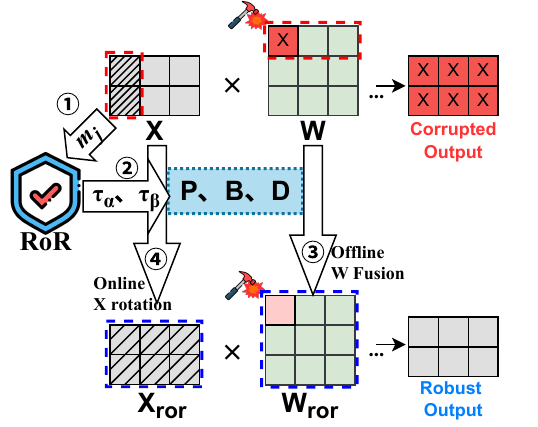}%
    }{%
        \fbox{\parbox[c][2.15in][c]{0.72\linewidth}{\centering
        Placeholder for revised Fig.~5\\[0.25em]
        $\mathbf{X}\rightarrow m_j\rightarrow(\tau_\alpha,\tau_\beta)$\\
        $\rightarrow\mathbf{P},\mathbf{B},\mathbf{D}$\\[0.15em]
        Offline: $\mathbf{W}_{\mathrm{RoR}}=\mathbf{Q}^{\top}\mathbf{W}$\\
        Online: $\mathbf{X}_{\mathrm{RoR}}=\mathbf{X}\mathbf{P}\mathbf{B}\mathbf{D}$}}
    }
    \caption{\rev{\textbf{Offline--online deployment of RoR.}
    Clean calibration profiles activation-channel sensitivity and constructs the compact $\mathbf{P}/\mathbf{B}/\mathbf{D}$ plan. Online inference applies $\mathbf{X}_{\mathrm{RoR}}=\mathbf{X}\mathbf{P}\mathbf{B}\mathbf{D}$, while the matched inverse transform $\mathbf{Q}^{\top}=\mathbf{D}\mathbf{B}^{\top}\mathbf{P}^{\top}$ is fused into the weights offline. For INT8 deployment, the fused weights are re-quantized after this transformation.}}
    \label{fig:ror_framework}
\end{figure}

{\ifmarked\color{blue}\fi
\noindent\textbf{Offline--online deployment.}
Fig.~\ref{fig:ror_framework} summarizes the RoR workflow. A clean calibration pass first profiles $m_j$ and constructs the sensitivity-aware plan, including $\mathbf{P}$, the hierarchy-consistent butterfly factor $\mathbf{B}$, and the seeded sign factor $\mathbf{D}$. The matching weight transform is fused once offline:
\begin{equation}
\label{eq:weight_fusion}
\begin{aligned}
    \mathbf{W}_{\mathrm{RoR}}
    &=\mathbf{Q}^{\top}\mathbf{W}\\
    &=\mathbf{D}\mathbf{B}^{\top}\mathbf{P}^{\top}\mathbf{W}.
\end{aligned}
\end{equation}
The dense matrix $\mathbf{Q}$ is never materialized; its structured factors are applied directly. For INT8 deployment, the fused weights are then re-quantized using the deployment quantization protocol. During inference,
\begin{equation}
\label{eq:act_transform}
    \mathbf{X}_{\mathrm{RoR}}
    =\mathbf{X}\mathbf{Q}
    =\mathbf{X}\mathbf{P}\mathbf{B}\mathbf{D},
    \qquad
    \mathbf{Y}
    =\mathbf{X}_{\mathrm{RoR}}\mathbf{W}_{\mathrm{RoR}}.
\end{equation}
Sensitivity profiling, plan construction, and weight fusion are one-time offline operations; only the structured activation-side transform is executed online. Backend-specific execution maps are derived once during model initialization and reused during inference.
}

{\ifmarked\color{blue}\fi
\noindent\textbf{Compute and latency overhead.}
For $N$ activation vectors, the baseline dense linear operation requires approximately
\begin{equation}
\label{eq:gemm_cost}
    C_{\mathrm{GEMM}}
    \simeq
    2N d_{in}d_{out}
\end{equation}
FLOPs. Let $d_j$ be the realized mixing depth of channel $j$. Since each local Householder operation jointly processes two features, the total number of active pairs per activation vector is
\begin{equation}
\label{eq:number_pairs}
    M=\frac{1}{2}\sum_{j=1}^{d_{in}}d_j.
\end{equation}
Under Eq.~\eqref{eq:pairwise_operation}, each pair requires two additions/subtractions and two constant multiplications. Including the sign factor, a conservative arithmetic count is
\begin{equation}
\label{eq:rotation_cost}
    C_{\mathrm{RoR}}
    \simeq
    4NM+Nd_{in}
    =2N\sum_j d_j+Nd_{in},
\end{equation}
and therefore
\begin{equation}
\label{eq:compute_overhead}
    \frac{C_{\mathrm{RoR}}}{C_{\mathrm{GEMM}}}
    \simeq
    \frac{\sum_j d_j+d_{in}/2}
         {d_{in}d_{out}}.
\end{equation}
The permutation $\mathbf{P}$ is a data-movement cost rather than a FLOP term and is therefore excluded from Eq.~\eqref{eq:compute_overhead} but included, together with memory traffic, kernel launches, and hardware utilization, in the measured end-to-end latency. For a $2048\times2048$ linear layer under full-depth mixing, the direct butterfly/sign arithmetic is approximately $0.56\%$ of the baseline GEMM arithmetic; measured Prefill and Decode latency is reported separately in Sec.~\ref{sec:evaluation}.
}

{\ifmarked\color{blue}\fi
\noindent\textbf{Persistent storage and runtime representation.}
The transformed weights $\mathbf{W}_{\mathrm{RoR}}$ replace the original deployment weights. RoR persistently stores only three compact vectors per protected linear layer: an INT32 permutation-index vector $\pi$ for $\mathbf{P}$, a UINT8 ranked realized-mixing-depth vector for $\mathbf{B}$, and an INT8 sign vector for $\mathbf{D}$. The sensitivity scores $m_j$ and requested mixing depths $r_j$ are offline planning quantities and are not required after plan construction.

The persistent RoR rotation parameters therefore require
\begin{equation}
\label{eq:storage_cost}
    S_{\mathrm{RoR}}
    \simeq
    (4+1+1)d_{in}
    =6d_{in}\ \text{bytes}
\end{equation}
per protected linear layer. When normalized against the INT8 qweight payload of a $d_{in}\times d_{out}$ layer,
\begin{equation}
\label{eq:storage_ratio}
    \frac{S_{\mathrm{RoR}}}{S_{\mathrm{W}}}
    \simeq
    \frac{6}{d_{out}}.
\end{equation}
For $d_{in}=d_{out}=2048$, this corresponds to $12$~KiB of persistent RoR rotation parameters for a $4$~MiB INT8 weight matrix, or approximately $0.29\%$. This analytical ratio uses the INT8 qweight payload alone as its denominator; the empirical storage overhead reported in Sec.~\ref{subsec:exp_efficiency} is normalized by the complete protected W8A16 persistent footprint, including FP32 per-output-channel scales. Backend-specific execution maps and temporary buffers are reconstructed or allocated at runtime and are not counted as persistent model storage.
}

{\ifmarked\color{blue}\fi
\begin{algorithm}[!t]
\caption{Rotated Robustness (RoR)}
\label{alg:ror}
\begin{algorithmic}[1]
\REQUIRE Pre-trained LLM weights $\{\mathbf{W}^{(l)}\}$, clean calibration dataset $\mathcal{D}_{cal}$, threshold parameters $\alpha$ and $\beta$, random seed $s$
\ENSURE RoR-transformed weights $\{\mathbf{W}_{\mathrm{RoR}}^{(l)}\}$ and compact $\mathbf{P}/\mathbf{B}/\mathbf{D}$ rotation parameters

\STATE \textbf{Phase 1: Channel-Sensitivity Profiling}
\FOR{each protected linear layer $l$}
    \STATE Profile $m_j^{(l)}=\max_i|x_{i,j}^{(l)}|$ over $\mathcal{D}_{cal}$ for all input channels
\ENDFOR

\STATE \textbf{Phase 2: RoR Plan Construction}
\FOR{each protected linear layer $l$}
    \STATE Compute $\mu^{(l)}$, $\sigma^{(l)}$, $\tau_{\alpha}^{(l)}$, and $\tau_{\beta}^{(l)}$
    \STATE Obtain descending ranking $\pi^{(l)}$ and permutation $\mathbf{P}^{(l)}$
    \STATE Compute requested mixing depths $r_j^{(l)}$ using Eq.~\eqref{eq:adaptive_depth}
    \STATE Construct the minimal hierarchy-consistent realized mixing depths $d_j^{(l)}\ge r_j^{(l)}$
    \STATE Define $\mathbf{B}^{(l)}$ from the ranked realized-depth vector using Eqs.~\eqref{eq:hierarchical_group}--\eqref{eq:b_levels}
    \STATE Generate the seeded sign factor $\mathbf{D}^{(l)}$
    \STATE Define $\mathbf{Q}^{(l)}\leftarrow\mathbf{P}^{(l)}\mathbf{B}^{(l)}\mathbf{D}^{(l)}$ implicitly

    \STATE \textbf{Phase 3: Offline Weight Fusion}
    \STATE Compute $\mathbf{W}_{\mathrm{RoR}}^{(l)}\leftarrow(\mathbf{Q}^{(l)})^{\top}\mathbf{W}^{(l)}$ using the structured factors
    \STATE Quantize $\mathbf{W}_{\mathrm{RoR}}^{(l)}$ for INT8 deployment
    \STATE Store $\mathbf{W}_{\mathrm{RoR}}^{(l)}$, $\pi^{(l)}$, realized mixing depths, and sign vector
\ENDFOR

\STATE \textbf{Phase 4: Online Inference}
\FOR{each protected linear layer $l$ during inference}
    \STATE $\mathbf{X}_{\mathrm{RoR}}^{(l)}\leftarrow\mathbf{X}^{(l)}\mathbf{P}^{(l)}\mathbf{B}^{(l)}\mathbf{D}^{(l)}$
    \STATE $\mathbf{Y}^{(l)}\leftarrow\mathbf{X}_{\mathrm{RoR}}^{(l)}\mathbf{W}_{\mathrm{RoR}}^{(l)}$
\ENDFOR
\end{algorithmic}
\end{algorithm}
}
\section{Evaluation}
\label{sec:evaluation}

\rev{We evaluate RoR within the Transformer Linear INT8 weight-bit domain defined in Sec.~\ref{subsec:threat_model}, covering random and gradient-guided bit-flip robustness, defense-aware SPoF reconstruction under white-box knowledge, downstream-task generalization, measured deployment overhead, and hyperparameter sensitivity.}

We organize our experiments to answer the following six research questions:
\begin{itemize}
\item \textbf{RQ1 (Black-Box Robustness):} How robust is RoR to random bit-flip attacks?

\item \textbf{RQ2 (Gray-Box Robustness):} How robust is RoR to gradient-guided Progressive Bit Search (PBS)?

\rev{\item \textbf{RQ3 (Defense-Aware SPoF Reconstruction):} Under white-box knowledge, what is the reconstruction cost of a verified Single Point of Failure (SPoF) within the Transformer Linear INT8 weight-bit domain?}

\item \textbf{RQ4 (Task Generalization):} Does RoR's robustness extend to downstream-task accuracy under targeted bit-flip attacks?

\rev{\item \textbf{RQ5 (Measured Efficiency):} What Prefill/Decode latency and persistent storage overhead does RoR incur?}

\rev{\item \textbf{RQ6 (Ablation Study):} How sensitive is RoR's PBS robustness to $\beta$, and does sensitivity-aware channel assignment improve robustness under a matched mixing budget?}

\end{itemize}

\subsection{Experimental Setup}
\label{subsec:setup}

\noindent\textbf{Implementation Details.}
We implement RoR in PyTorch with CUDA support. All experiments are conducted on a server equipped with an Intel Xeon Platinum 8558P CPU and a single NVIDIA H200 NVL GPU. The defense and attack implementations share the same quantization and evaluation infrastructure unless otherwise specified in the corresponding research question.

\rev{\noindent\textbf{Models.}
We evaluate eight Transformer language models spanning multiple scales and model families: OPT-125M~\cite{zhang2022opt}, SmolLM2-360M~\cite{allal2025smollm2}, Qwen2.5-0.5B~\cite{team2024qwen2}, Llama-3.2-1B~\cite{dubey2024llama}, Llama-2-7B~\cite{touvron2023llama}, Qwen2.5-7B, Qwen2.5-14B, and Mistral-Nemo-Base-2407. The specific model subsets used by individual research questions are stated in the corresponding subsections.}

\noindent\textbf{Datasets.}
We use WikiText-2~\cite{merity2016pointer} as the primary language-modeling benchmark and report perplexity (PPL) to quantify model degradation under bit-flip attacks. \rev{For WikiText-2, RoR profiles activation-channel sensitivity using 128 sequences of 2,048 tokens from the training split, while PPL is evaluated on fixed 2,048-token windows drawn from the official test split. We use PPL $>100$ as a common operational catastrophic-failure threshold because it exceeds the clean PPL range of all evaluated deployments and therefore provides a consistent criterion for identifying severe language-modeling degradation beyond ordinary clean or finite-precision variation. This threshold is used only as a common reporting criterion; continuous PPL values are reported throughout the evaluation so that the conclusions do not depend solely on this cutoff. For downstream-task generalization, we evaluate HellaSwag~\cite{zellers2019hellaswag}, PIQA~\cite{bisk2020piqa}, and a four-domain MMLU subset~\cite{hendrycks2020measuring} consisting of elementary mathematics, computer security, philosophy, and global facts. RQ4 evaluates these tasks in a 5-shot setting.}

{\ifmarked\color{blue}\fi
\noindent\textbf{Numerical Format and Quantization Protocol.}
All random bit-flip, PBS, and SPoF-reconstruction experiments operate on per-output-channel symmetrically quantized INT8 two's-complement Transformer Linear weights with FP32 scales. INT8 bit positions are indexed from 0 (LSB) to 7 (MSB/sign bit). The writable fault domain is restricted to these Linear weight bits and excludes embeddings, normalization parameters, quantization scales, and defense metadata/control state, as defined in Sec.~\ref{subsec:threat_model}. All transformed weights are quantized after the corresponding defense transformation; therefore, clean performance refers to the deployed quantized model rather than assuming strict losslessness under finite precision. In RQ3, SPoF reconstruction cost is measured as the XOR Hamming distance between the clean deployed INT8 weight representation and its reconstructed attacked counterpart. In RQ5, persistent defense-storage overhead is normalized by the protected W8A16 baseline persistent footprint, consisting of the INT8 Linear weights and FP32 per-output-channel scales.
}

\rev{\noindent\textbf{Compared Deployments.}}
We compare RoR with the following representative undefended, structural, detection-based, and rotation-based deployments. The key configurations below are fixed across the corresponding robustness experiments unless otherwise stated.

\textit{(i) Baseline}: The vanilla W8A16 quantized model with no defense applied, serving as the unprotected reference.

\textit{(ii) FaR~\cite{nazari2024forget}}: A training-free structural defense that rewires vulnerable Transformer Linear connections. \rev{We use $p{=}5{\times}10^{-4}$ and division factor $D{=}5$. The relatively small rewiring ratio is chosen for LLMs to limit the number of modified connections and the associated clean-performance and deployment overhead, while retaining FaR's rewiring mechanism and redundancy strength.}

\textit{(iii) RADAR~\cite{li2021radar}}: A runtime detection-and-recovery defense based on group-level weight signatures. \rev{We use group size $G{=}512$, following the configuration evaluated in the original RADAR study, and retain its signature and recovery semantics.}

\rev{\textit{(iv) QuaRot-Had~\cite{ashkboos2024quarot}}: QuaRot was originally proposed to improve low-bit LLM quantization through orthogonal rotation and was not designed as a defense against bit-flip attacks. We implement QuaRot-Had by following the Hadamard rotation and deployment pattern in the official QuaRot codebase. Because it shares the use of orthogonal activation/weight transformations with RoR, we include it as a rotation-based methodological reference under the same quantized BFA evaluation protocol.}

\textit{(v) RoR}: Our sensitivity-aware hierarchical rotation using the layer-wise $\mathbf{Q}=\mathbf{P}\mathbf{B}\mathbf{D}$ construction in Sec.~\ref{sec:method}. \rev{We fix $\alpha{=}6$ and use $\beta{=}0.15$ for sub-1B models and $\beta{=}0.30$ for models with at least 1B parameters. RQ6 evaluates the sensitivity to $\beta$ and the contribution of sensitivity-aware channel assignment.}

\subsection{Black-Box Robustness under Random Bit-Flip Attacks (RQ1)}
\label{subsec:rq1}

{\ifmarked\color{blue}\fi
Under the black-box setting, we perform 1,000 BER-based Monte Carlo trials on Qwen2.5-0.5B, Llama-3.2-1B, Llama-2-7B, and Qwen2.5-7B using a unified BER of $3{\times}10^{-4}$. In each trial, every eligible INT8 bit of the protected Transformer Linear weights is independently sampled for flipping. For reproducibility and paired comparison, all evaluated deployments use the same 1,000 fixed random seeds, ranging from 4000 to 4999, for each evaluated model. The realized flip counts are $858{,}718{\pm}914$, $2{,}335{,}297{\pm}1{,}507$, $15{,}542{,}041{\pm}3{,}942$, and $15{,}660{,}334{\pm}3{,}956$ (mean $\pm$ standard deviation) for the four models, respectively. Thus, RQ1 represents a dense stochastic BER stress protocol, distinct from the sparse targeted-fault settings in RQ2 and RQ3. We evaluate WikiText-2 perplexity (PPL) and define a failure as PPL $>100$.

\noindent\textbf{Results.}
Table~\ref{tab:random_flip_combined} summarizes the maximum observed PPL and empirical failure rate over the 1,000 Monte Carlo trials.

\begin{table*}[t]
    \centering
    \caption{Robustness under random bit-flip attacks with BER $3{\times}10^{-4}$ over 1,000 Monte Carlo trials. A failure is defined as PPL $>100$.}
    \label{tab:random_flip_combined}
    \scriptsize
    \resizebox{\textwidth}{!}{%
    \begin{tabular}{lcccccccc}
        \toprule
        \multirow{2}{*}{\textbf{Method}}
        & \multicolumn{2}{c}{\textbf{Qwen2.5-0.5B}}
        & \multicolumn{2}{c}{\textbf{Llama-3.2-1B}}
        & \multicolumn{2}{c}{\textbf{Llama-2-7B}}
        & \multicolumn{2}{c}{\textbf{Qwen2.5-7B}} \\
        \cmidrule(lr){2-3}
        \cmidrule(lr){4-5}
        \cmidrule(lr){6-7}
        \cmidrule(lr){8-9}
        & \textbf{Max PPL} & \textbf{Fail. Rate}
        & \textbf{Max PPL} & \textbf{Fail. Rate}
        & \textbf{Max PPL} & \textbf{Fail. Rate}
        & \textbf{Max PPL} & \textbf{Fail. Rate} \\
        \midrule
        Baseline
        & 225.89 & 0.2\%
        & 152.54 & 0.1\%
        & 933.57 & 0.1\%
        & $3.25{\times}10^{4}$ & 3.4\% \\
        FaR
        & 30.41 & 0.0\%
        & 20.38 & 0.0\%
        & 9.24 & 0.0\%
        & 11.38 & 0.0\% \\
        RADAR
        & $7.05{\times}10^{7}$ & 100.0\%
        & $2.45{\times}10^{6}$ & 100.0\%
        & $1.05{\times}10^{5}$ & 100.0\%
        & $2.99{\times}10^{5}$ & 100.0\% \\
        QuaRot-Had
        & 15.49 & 0.0\%
        & 12.21 & 0.0\%
        & 6.35 & 0.0\%
        & 8.39 & 0.0\% \\
        \textbf{RoR (Ours)}
        & 15.52 & 0.0\%
        & 12.23 & 0.0\%
        & 6.36 & 0.0\%
        & 8.43 & 0.0\% \\
        \bottomrule
    \end{tabular}%
    }
\end{table*}

The undefended models exhibit architecture-dependent catastrophic tails, most prominently on Qwen2.5-7B, where the failure rate reaches 3.4\% and the maximum PPL reaches $3.25{\times}10^{4}$. No FaR failures are observed on the four evaluated models, although FaR retains higher worst-case PPL than the rotation-based deployments. RADAR performs poorly under this dense BER protocol because faults span many protected groups and its recovery suppresses detected corrupted groups rather than reconstructing their original weights; this result therefore characterizes RADAR under dense stochastic corruption rather than its general detection capability. Both rotation-based deployments exhibit no observed failures: RoR limits the maximum PPL to 15.52, 12.23, 6.36, and 8.43 across the four models, with QuaRot-Had showing similarly compact tails. These zero observed failures are empirical results under the tested 1,000-trial protocol and should not be interpreted as a zero-risk guarantee.

}
\subsection{Gray-Box Robustness under Gradient-Guided PBS Attacks (RQ2)}
\label{subsec:rq2_pbs}

{\ifmarked\color{blue}\fi
Under the gray-box setting, the adversary has access to each method's deployed quantized Linear weights and gradients with respect to those weights, but not the internal defense configuration. We evaluate gradient-guided Progressive Bit Search (PBS)~\cite{rakin2019bit} on all eight models, restricting flips to eligible INT8 bits in Transformer Linear weights. PBS is independently re-optimized for each deployed method rather than transferring a bit sequence found on the undefended Baseline, and gradients are recomputed on the current quantized model after every accumulated flip. We report the first cumulative PBS bit count at which WikiText-2 PPL exceeds 100 as a compact measure of sustained attack robustness; complete unclipped results are provided in the Supplemental Material.

\begin{figure*}[t]
    \centering
    \includegraphics[width=0.98\textwidth]{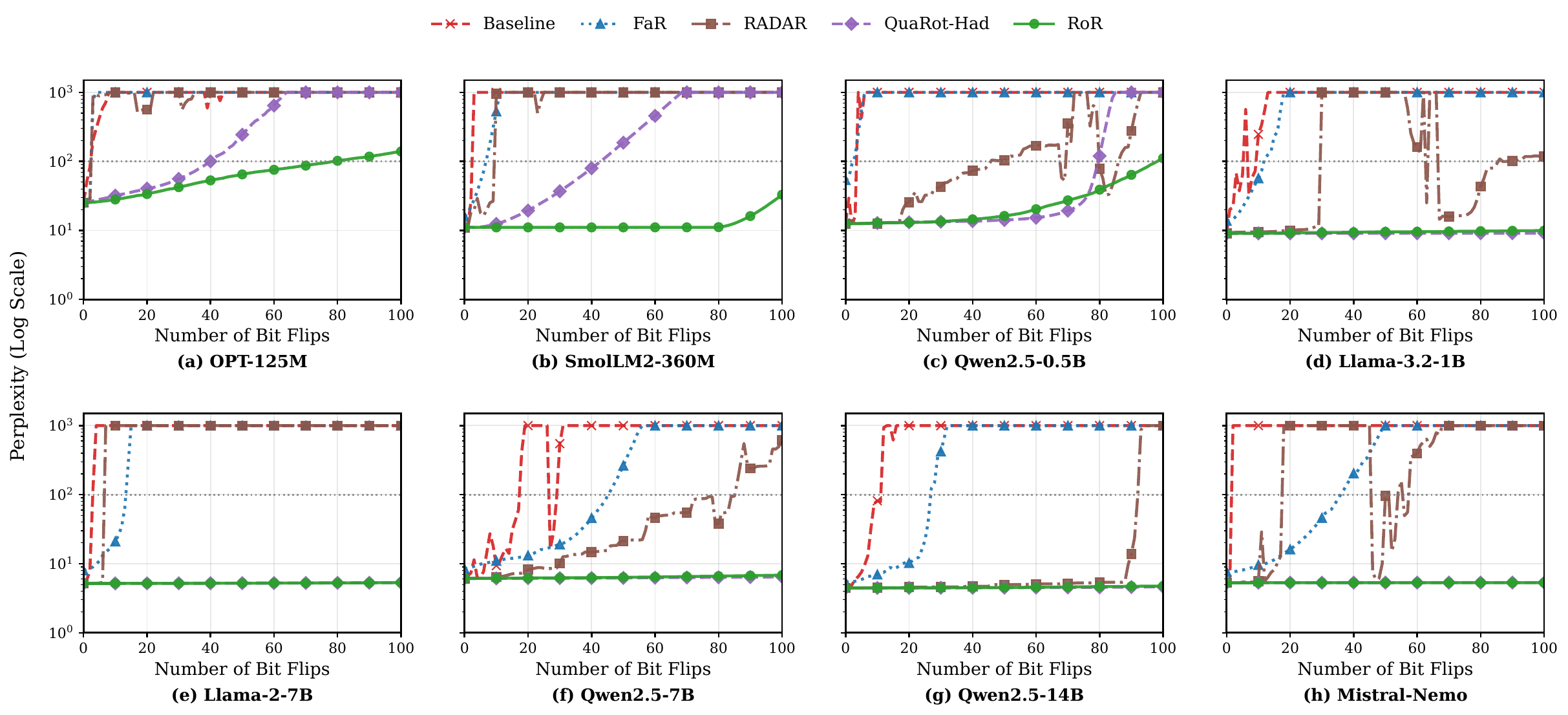}
    \caption{PPL degradation under gray-box gradient-guided PBS attacks across eight models. Each curve corresponds to one deployment method. PPL values above 1000 are clipped for readability, and the horizontal threshold denotes PPL$=100$.}
    \label{fig:rq2_pbs}
\end{figure*}

\noindent\textbf{Results.}
Figure~\ref{fig:rq2_pbs} shows a clear separation between the non-rotation defenses and the two rotation-based deployments. The undefended Baseline crosses PPL$=100$ within 2--18 cumulative flips across all eight models. FaR improves the threshold-crossing budget on six models but still exceeds the threshold within 46 flips in every case. RADAR provides stronger early protection on several models, yet all eight deployments cross PPL$=100$ within PBS100; its non-monotonic trajectories are consistent with group-level detection and recovery, where additional flips can change which corrupted groups are detected and suppressed.

RoR achieves the largest or tied-largest threshold-crossing budget on all eight models. Compared with QuaRot-Had, RoR extends the budget from 40 to 79 flips on OPT-125M, from 43 to beyond PBS100 on SmolLM2-360M, and from 80 to 99 flips on Qwen2.5-0.5B. On the remaining five models, neither rotation-based deployment exceeds PPL$=100$ within PBS100 and their trajectories remain close. Thus, RoR improves sustained PBS robustness over QuaRot-Had on the compact models where the generic rotation reference eventually fails, while matching its strong robustness on the other evaluated architectures.

}
{\ifmarked\color{blue}\fi
\subsection{Defense-Aware SPoF Reconstruction within the Linear Weight-Bit Domain (RQ3)}
\label{subsec:rq3}

Under the white-box knowledge setting, the adversary is assumed to know the model weights, gradients, complete defense algorithm, and deployment configuration. However, consistent with the threat domain defined in Section~\ref{sec:background_threat}, the writable fault domain remains restricted to eligible INT8 two's-complement weight bits in Transformer Linear layers. RQ3 therefore does not evaluate arbitrary corruption of embeddings, normalization parameters, quantization scales, or defense metadata. Instead, it asks a narrower defense-aware question: given a verified single-bit SPoF in the undefended INT8 model, how many deployed INT8 weight-bit changes are required to reconstruct the specified SPoF perturbation and reproduce catastrophic failure in each defended deployment?

We use three manually verified baseline SPoFs. On OPT-125M, the source fault is bit~7 of \texttt{layer2.fc1}$[643,706]$, which raises the WikiText-2 PPL from 25.17 to $2.48{\times}10^{3}$. On Qwen2.5-7B, the source fault is bit~7 of \texttt{layer1.mlp.gate\_proj}$[17094,1923]$, which raises the PPL from 6.13 to $4.28{\times}10^{5}$. On Llama-2-7B, the source fault is bit~7 of \texttt{layer1.self\_attn.v\_proj}$[1100,1512]$, which raises the PPL from 5.18 to $2.18{\times}10^{4}$. Under PyTorch's $[\text{out\_features},\text{in\_features}]$ Linear-weight convention, the first index denotes the output row and the second denotes the input-channel column.

\noindent\textbf{SPoF--Sensitivity Alignment.}
Before reconstructing these source perturbations, we examine whether their corresponding input channels align with the activation sensitivity that motivates RoR. We use the same calibration-time sensitivity metric defined in Section~\ref{sec:method},
$m_j=\|\mathbf{X}_{:,j}\|_{\infty}$, and rank each SPoF-coupled channel among all input channels of the corresponding Linear layer. As shown in Table~\ref{tab:spof_sensitivity}, all three verified SPoFs are associated with channels in the top 0.3\% of their layer-wise sensitivity rankings and all exceed the high-sensitivity threshold $\tau_\alpha$. In particular, the Llama-2-7B and Qwen2.5-7B SPoFs are both coupled to the single most sensitive input channel in their respective layers. These cases provide cross-model evidence consistent with the fault-amplification mechanism motivating RoR.

\begin{table}[t]
\centering
\caption{Alignment of verified SPoFs with high-sensitivity activation channels. Rank is computed among all input channels of the corresponding Linear layer using $m_j=\|\mathbf{X}_{:,j}\|_{\infty}$, where rank~1 denotes the highest sensitivity.}
\label{tab:spof_sensitivity}
\scriptsize
\resizebox{\columnwidth}{!}{%
\begin{tabular}{lcccc}
\toprule
\textbf{Model}
& \textbf{$m_j$}
& \textbf{Rank / $d_{\mathrm{in}}$}
& \textbf{Top \%}
& \textbf{$m_j/\tau_\alpha$} \\
\midrule
OPT-125M
& 7.81
& 2 / 768
& 0.260\%
& 1.43$\times$ \\
Llama-2-7B
& 9.44
& 1 / 4096
& 0.0244\%
& 3.91$\times$ \\
Qwen2.5-7B
& 141.00
& 1 / 3584
& 0.0279\%
& 6.01$\times$ \\
\bottomrule
\end{tabular}%
}
\end{table}

\noindent\textbf{Attack Construction.}
Each defense induces a different deployed weight representation. We therefore reconstruct the verified baseline SPoF according to the corresponding defense mechanism and count the resulting INT8 bit changes after quantization.

\textit{(i) Baseline} directly replays the verified single-bit SPoF at the original source location.

\textit{(ii) FaR} reconstructs the SPoF through its known rewiring configuration by mapping the source perturbation to the corresponding rewired weight entries.

\textit{(iii) RADAR} reconstructs the SPoF through a checksum-preserving collision by adding a compensation bit within the same protected group so that the live checksum remains unchanged while the SPoF-inducing perturbation is replayed.

\textit{(iv) Rotation-based deployments} reconstruct the same source SPoF in their respective rotated weight representations. RoR uses the deployed sensitivity-aware hierarchical rotation $\mathbf{Q}_{\mathrm{RoR}}=\mathbf{P}\mathbf{B}\mathbf{D}$, while QuaRot-Had uses the randomized generalized Hadamard rotation $\mathbf{Q}_{\mathrm{Had}}=\mathbf{H}\mathbf{D}$. The reconstructed floating-point perturbation is then quantized into the deployed INT8 representation, and the resulting bit changes are counted. This procedure does not establish a global lower bound for all possible successful attacks; rather, it measures how strongly the specified baseline SPoF is delocalized by each deployed representation.

\noindent\textbf{Results.}
Table~\ref{tab:rq3_whitebox} reports the SPoF reconstruction cost, number of touched weights, and replay PPL for the three verified SPoFs. SPoF reconstruction cost is defined as the actual INT8 bit changes measured by the XOR Hamming distance between the clean deployed INT8 weight representation and its reconstructed attacked counterpart.

\textit{(i) FaR and RADAR require only a small number of coordinated bit changes.}
Across the three verified SPoFs, FaR requires one to six coordinated INT8 bit changes, while RADAR requires two through a checksum-preserving collision; the undefended Baseline requires only the original single-bit fault. Thus, for these specified SPoFs, the two non-rotation defenses increase reconstruction cost only modestly.

\textit{(ii) Rotation-based deployments substantially increase SPoF reconstruction cost.}
RoR requires 1,906, 6,133, and 12,197 INT8 bit changes on OPT-125M, Qwen2.5-7B, and Llama-2-7B, respectively, while QuaRot-Had requires 2,714, 12,341, and 12,264 bit changes. These changes are distributed across 512--4,095 deployed weights for RoR and 767--4,095 for QuaRot-Had. The results should not be interpreted as a universal lower bound on all possible defense-aware attacks; instead, they quantify the cost of reconstructing the specified baseline SPoFs within the stated Transformer Linear INT8 weight-bit domain. The roughly three- to four-order-of-magnitude increase relative to the original one-bit faults shows that rotation-based deployments strongly delocalize the specified SPoFs.

\begin{table}[t]
\centering
\caption{Defense-aware SPoF reconstruction on OPT-125M, Qwen2.5-7B, and Llama-2-7B. Reconstruction cost is the INT8 XOR Hamming distance between the clean deployed weight representation and its reconstructed attacked counterpart.}
\label{tab:rq3_whitebox}
\scriptsize
\resizebox{\columnwidth}{!}{%
\begin{tabular}{llccc}
\toprule
\textbf{Model}
    & \textbf{Method}
    & \textbf{SPoF Recon. Bits}
    & \textbf{Weights Touched}
    & \textbf{Replay PPL} \\
\midrule
\multirow{5}{*}{OPT-125M}
    & Baseline
    & 1 & 1 & $2.48{\times}10^{3}$ \\
    & FaR
    & 1 & 1 & $2.84{\times}10^{3}$ \\
    & RADAR
    & 2 & 2 & $2.47{\times}10^{3}$ \\
    & QuaRot-Had
    & 2,714 & 767 & $1.91{\times}10^{3}$ \\
    & \textbf{RoR (Ours)}
    & 1,906 & 512 & $2.87{\times}10^{3}$ \\
\midrule
\multirow{5}{*}{Qwen2.5-7B}
    & Baseline
    & 1 & 1 & $4.28{\times}10^{5}$ \\
    & FaR
    & 6 & 6 & $6.43{\times}10^{5}$ \\
    & RADAR
    & 2 & 2 & $4.29{\times}10^{5}$ \\
    & QuaRot-Had
    & 12,341 & 3,583 & $5.36{\times}10^{5}$ \\
    & \textbf{RoR (Ours)}
    & 6,133 & 2,048 & $2.04{\times}10^{5}$ \\
\midrule
\multirow{5}{*}{Llama-2-7B}
    & Baseline
    & 1 & 1 & $2.18{\times}10^{4}$ \\
    & FaR
    & 6 & 6 & $1.78{\times}10^{4}$ \\
    & RADAR
    & 2 & 2 & $2.18{\times}10^{4}$ \\
    & QuaRot-Had
    & 12,264 & 4,095 & $1.77{\times}10^{4}$ \\
    & \textbf{RoR (Ours)}
    & 12,197 & 4,095 & $2.94{\times}10^{4}$ \\
\bottomrule
\end{tabular}%
}
\end{table}

}
\subsection{Task Generalization under Attack (RQ4)}
\label{subsec:exp_generalization}

To determine whether RoR's robustness extends beyond WikiText-2 perplexity to downstream-task utility, we replay the cumulative targeted PBS bit-flip sequences obtained in RQ2 and evaluate task accuracy on three downstream benchmarks.

{\ifmarked\color{blue}\fi
\noindent\textbf{Benchmarks and Protocol.}
We evaluate a four-domain MMLU subset~\cite{hendrycks2020measuring}, HellaSwag~\cite{zellers2019hellaswag}, and PIQA~\cite{bisk2020piqa}. The MMLU subset consists of Elementary Mathematics, Computer Security, Philosophy, and Global Facts, and we report the macro-average accuracy across the four domains rather than the full MMLU benchmark. We use a 5-shot setting ($k{=}5$), with 100 evaluated examples per MMLU domain (400 in total) and 100 examples each for HellaSwag and PIQA. For each deployment method, we replay its corresponding cumulative PBS sequence from the clean deployment to 100 bit flips without re-running the attack search on the downstream tasks. Two-sided 95\% Wilson confidence intervals are used to characterize finite-sample uncertainty in task accuracy. The main text reports Qwen2.5-7B as a representative case; PBS100 endpoint results for all three evaluated models, including 95\% confidence intervals, are provided in Supplemental Material.
}

\begin{figure*}[t]
   \centering
   \includegraphics[width=0.95\textwidth]{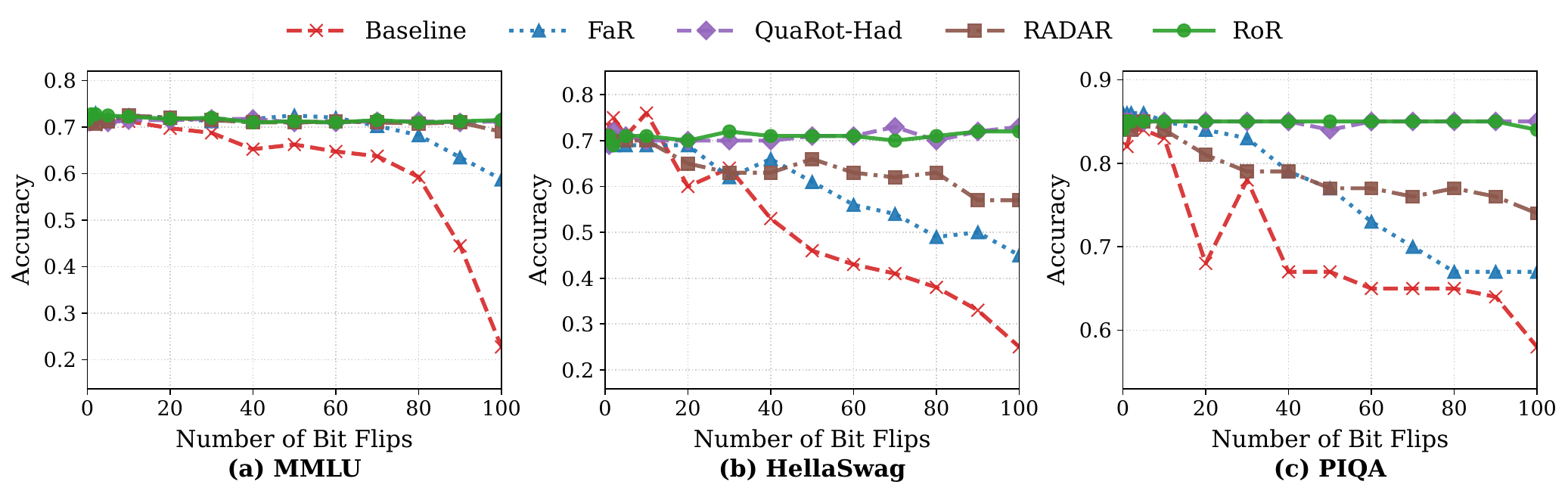}
   \caption{\rev{Task generalization under cumulative PBS attacks on Qwen2.5-7B. Accuracy on the four-domain MMLU subset, HellaSwag, and PIQA is evaluated from the clean deployment to 100 cumulative bit flips. The comparison includes Baseline, FaR, RADAR, QuaRot-Had, and RoR.}}
   \label{fig:generalization_degradation}
\end{figure*}

{\ifmarked\color{blue}\fi
\noindent\textbf{Results.}
Figure~\ref{fig:generalization_degradation} shows that downstream-task accuracy follows the same qualitative trend as the WikiText-2 PPL results. Baseline degrades sharply under cumulative PBS, while FaR and RADAR retain partial task utility. In contrast, both rotation-based deployments remain close to their clean performance through PBS100: RoR achieves 71.50\% on the four-domain MMLU subset, 72.0\% on HellaSwag, and 84.0\% on PIQA, while QuaRot-Had reaches 71.25\%, 73.0\%, and 85.0\%. These results show that RoR's robustness extends beyond language-modeling perplexity to downstream-task utility. Small endpoint fluctuations around clean accuracy fall within the finite-sample uncertainty of the fixed evaluation subsets and are not interpreted as attack-induced improvements.
}

{\ifmarked\color{blue}\fi
\subsection{Inference Efficiency \& Overhead (RQ5)}
\label{subsec:exp_efficiency}

We evaluate deployment overhead under a unified W8A16 CUDA-Graph inference harness on an NVIDIA H200. All defense-specific offline preparation is completed before timing, and all timed Transformer Linear operations use the same fixed-dispatch Triton W8A16 GEMM backend with contiguous INT8 weights, per-output-channel FP32 scales, and FP16/BF16 activations. We report four representative models spanning different scales and families: OPT-125M, Llama-3.2-1B, Qwen2.5-7B, and Mistral-Nemo-Base-2407. All measurements use batch size 8; Prefill uses sequence length 2048, while Decode measures one-token autoregressive latency after a 2048-token context, except for OPT-125M, which uses 2047 tokens because of its 2048-position limit. Each latency is averaged over five timed CUDA-Graph replays after two warm-up replays.

\noindent\textbf{Implementation and Accounting Scope.}
All methods use the same StaticCache CUDA-Graph protocol and common W8A16 Linear backend. FaR is implemented with a GPU realization that preserves the released rewiring semantics, with frozen rewiring rules executed through a graph-capturable CUDA correction path on top of the common W8A16 Linear output. RADAR uses group size $G{=}512$ and computes group signatures directly from the resident INT8 weights before the common GEMM; the reported fault-free latency includes checksum computation but excludes recovery, which is invoked only after detection. QuaRot-Had and RoR both perform their matched weight transformations offline and apply their corresponding activation transforms online through CUDA implementations.

The FaR and RADAR measurements should not be interpreted as direct reproductions of the runtime overheads reported in their original papers because their execution substrates and workloads differ substantially from our LLM deployment setting. FaR was originally evaluated on substantially smaller Transformer workloads using RTX 3070 and Jetson Nano platforms, and its paper notes that the customized Linear path lacks optimized low-level library support and that further optimization is needed for multi-billion-parameter LLMs~\cite{nazari2024forget}. Our results instead measure the cost of preserving FaR's rewiring semantics within a unified H200 W8A16 autoregressive inference path. RADAR's published timing evaluation uses a gem5-based Arm CPU system model with ResNet workloads, whereas its released implementation provides the PyTorch detection/recovery logic rather than an H200 autoregressive deployment~\cite{li2021radar}. We therefore realize the same group-signature semantics as a software checksum over resident INT8 weights. This cost is more readily amortized during Prefill, where large GEMMs dominate execution, but becomes comparatively more visible during one-token Decode. Accordingly, the FaR and RADAR results below should be interpreted as unified GPU deployment costs under our common LLM harness rather than reproductions of the platform-specific overheads reported in the original works.

Persistent-storage accounting includes the protected W8A16 model footprint and defense-specific resident metadata. The Baseline consists of INT8 Linear-weight payloads and FP32 per-output-channel scales, and all storage overheads are normalized to this footprint. Temporary runtime buffers, StaticCache tensors, CUDA-Graph private pools, calibration activations, and offline artifacts are excluded. FaR stores frozen INT32 row/target/dead-neuron indices; RADAR stores a 16-bit key per protected Linear layer and packed 2-bit-per-group golden signatures; QuaRot-Had stores the persistent INT8 Rademacher sign vector $\mathbf{D}$; and RoR stores the per-layer INT32 permutation, INT8 sign vector, and UINT8 realized mixing-depth vector.

\noindent\textbf{Results.}
Table~\ref{tab:efficiency_comparison} reports the measured deployment cost. RoR incurs 5.2\%--14.1\% Prefill overhead and 0.5\%--9.8\% Decode overhead across the four models, consistently reducing Prefill cost relative to QuaRot-Had while providing lower or comparable Decode overhead. Under the same GPU harness, FaR exhibits substantially larger software overhead, whereas RADAR maintains low Prefill overhead but shows strongly model-dependent Decode cost. RoR requires 0.097\%--0.584\% additional persistent storage, trading a small metadata footprint for comparatively uniform runtime overhead across model scales and architectures.

\begin{table*}[t]
\centering
\caption{Measured H200 deployment cost. Each cell reports Storage / Prefill / Decode. Baseline values are absolute (GiB / ms / ms-token$^{-1}$), while defense rows report percentage overhead relative to the corresponding Baseline.}
\label{tab:efficiency_comparison}
\scriptsize
\resizebox{\textwidth}{!}{%
\begin{tabular}{lcccc}
\toprule
\textbf{Method}
& \textbf{OPT-125M}
& \textbf{Llama-3.2-1B}
& \textbf{Qwen2.5-7B}
& \textbf{Mistral-Nemo-Base-2407} \\
\midrule
Baseline
& 0.0794 / 32.93 / 4.33
& 0.9077 / 202.20 / 13.85
& 6.0823 / 912.48 / 35.73
& 10.1630 / 1524.59 / 55.39 \\
FaR
& +1.395\% / +148.5\% / +51.3\%
& +1.398\% / +333.7\% / +92.2\%
& +1.399\% / +433.6\% / +117.8\%
& +1.399\% / +759.1\% / +220.7\% \\
RADAR
& +0.049\% / +2.2\% / $-1.4$\%
& +0.049\% / +3.4\% / +31.9\%
& +0.049\% / +4.4\% / +123.0\%
& +0.049\% / +4.9\% / +130.8\% \\
QuaRot-Had
& +0.097\% / +68.2\% / +15.1\%
& +0.034\% / +13.9\% / +7.1\%
& +0.017\% / +22.7\% / +5.6\%
& +0.016\% / +15.3\% / +4.1\% \\
\textbf{RoR (Ours)}
& +0.584\% / +14.1\% / +9.8\%
& +0.202\% / +9.4\% / +0.5\%
& +0.104\% / +6.1\% / +4.8\%
& +0.097\% / +5.2\% / +4.1\% \\
\bottomrule
\end{tabular}%
}
\end{table*}

}
{\ifmarked\color{blue}\fi
\subsection{Ablation Study: Sensitivity to $\beta$ and Channel Assignment (RQ6)}
\label{subsec:exp_ablation}

We examine the sensitivity of RoR to $\beta$, which controls the lower sensitivity boundary $\tau_{\beta}=\beta\tau_{\alpha}$ in the graded mixing-depth assignment. We evaluate $\beta\in\{0.15,0.30,0.45\}$ with independent PBS runs up to 100 cumulative bit flips. Smaller $\beta$ generally assigns deeper mixing to more channels, whereas larger $\beta$ yields a more selective plan with lower online butterfly workload. Table~\ref{tab:ablation} reports WikiText-2 PPL after 30, 50, 75, and 100 cumulative flips.

\begin{table}[t]
\centering
\caption{Sensitivity of RoR to $\beta$. PPL@K denotes WikiText-2 perplexity after $K$ cumulative PBS flips. Values of $\beta$ used in the main experiments are bolded.}
\label{tab:ablation}
\scriptsize
\resizebox{\columnwidth}{!}{%
\begin{tabular}{llcccc}
\toprule
\textbf{Model}
& \textbf{$\beta$}
& \textbf{PPL@30}
& \textbf{PPL@50}
& \textbf{PPL@75}
& \textbf{PPL@100} \\
\midrule
\multirow{3}{*}{OPT-125M}
& \textbf{0.15} & 42.35 & 64.84 & 93.92 & 139.35 \\
& 0.30 & 57.06 & 94.26 & $4.48{\times}10^{3}$ & $5.07{\times}10^{3}$ \\
& 0.45 & 57.83 & 95.24 & $4.53{\times}10^{3}$ & $4.99{\times}10^{3}$ \\
\midrule
\multirow{3}{*}{SmolLM2-360M}
& \textbf{0.15} & 11.07 & 11.07 & 11.09 & 32.97 \\
& 0.30 & 96.80 & $1.35{\times}10^{3}$ & $5.53{\times}10^{4}$ & $3.38{\times}10^{6}$ \\
& 0.45 & 135.95 & $2.58{\times}10^{3}$ & $2.31{\times}10^{5}$ & $2.89{\times}10^{7}$ \\
\midrule
\multirow{3}{*}{Llama-2-7B}
& 0.15 & 5.21 & 5.23 & 5.28 & 5.34 \\
& \textbf{0.30} & 5.20 & 5.23 & 5.26 & 5.31 \\
& 0.45 & 5.20 & 5.21 & 5.25 & 5.29 \\
\midrule
\multirow{3}{*}{Qwen2.5-7B}
& 0.15 & 6.19 & 6.32 & 6.56 & 7.71 \\
& \textbf{0.30} & 6.25 & 6.35 & 6.57 & 6.84 \\
& 0.45 & 6.22 & 6.44 & 6.87 & 9.14 \\
\bottomrule
\end{tabular}%
}
\end{table}

\noindent\textbf{Robustness--Mixing Trade-off.}
The two compact models benefit substantially from broader/deeper mixing: $\beta{=}0.15$ gives the best sustained robustness on OPT-125M and SmolLM2-360M, while larger values eventually lead to sharp PPL growth. In contrast, both 7B models remain robust across the tested range; $\beta{=}0.30$ preserves comparable robustness with a more selective mixing plan, whereas $\beta{=}0.45$ begins to degrade more noticeably on Qwen2.5-7B. We therefore use $\beta{=}0.15$ for sub-1B models and $\beta{=}0.30$ for models with at least 1B parameters in the main experiments, avoiding broader mixing when a more selective plan is sufficient.

\noindent\textbf{Effect of Sensitivity-Aware Channel Assignment.}
To isolate the contribution of sensitivity-aware ordering, we construct a matched-cost random-assignment control by replacing only the sensitivity-sorted permutation $\mathbf{P}_{\mathrm{sens}}$ with a random permutation $\mathbf{P}_{\mathrm{rand}}$, while keeping the butterfly transform $\mathbf{B}$ and sign transform $\mathbf{D}$ unchanged. The two variants therefore share the same butterfly structure, mixing-depth distribution, and online workload, differing only in how original channels are assigned to the available mixing depths.

\begin{figure}[t]
    \centering
    \includegraphics[width=\columnwidth]{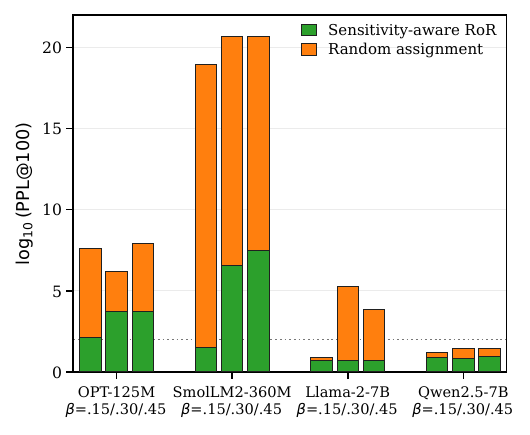}
    \caption{Matched-cost ablation of sensitivity-aware channel assignment. Bars show $\log_{10}(\mathrm{PPL@100})$ for RoR and Random Assignment across three $\beta$ values; lower is better.}
    \label{fig:rq6_matched_cost}
\end{figure}

Figure~\ref{fig:rq6_matched_cost} shows that sensitivity-aware RoR achieves lower PPL@100 than Random Assignment in all 12 model--$\beta$ configurations. The gap is especially large at the main configurations: 32.97 versus $9.08{\times}10^{18}$ on SmolLM2-360M at $\beta{=}0.15$, and 5.31 versus $1.81{\times}10^{5}$ on Llama-2-7B at $\beta{=}0.30$. Clean PPL remains nearly unchanged between the two assignments, indicating that the robustness gain arises from sensitivity-aware allocation of the same mixing budget rather than a clean-performance difference.

}

\section{Discussion and Future Work}
\label{sec:discussion}

RoR provides strong bit-flip robustness with low measured deployment overhead on the evaluated text-based LLMs, while several limitations and optimization opportunities remain. We discuss runtime optimization, protection beyond dense Linear layers, and extension to multimodal models.

\subsection{Further Reducing Deployment Overhead}

\rev{Although the structured RoR transform has low arithmetic cost relative to the protected GEMM, its measured latency also reflects activation movement, kernel scheduling, and hardware utilization. Further optimization can therefore focus on fusing the structured $\mathbf{P}/\mathbf{B}/\mathbf{D}$ activation transform with adjacent GEMM kernels and reducing memory-movement and launch overhead, particularly for repeated Decode operations.}

\subsection{Scope Limitation: Normalization Layers}

\rev{RoR currently protects dense Transformer Linear layers rather than normalization parameters.} For RMSNorm/LayerNorm, orthogonal rotation preserves the input norm, i.e., $\|\mathbf{xQ}\|_2=\|\mathbf{x}\|_2$, but the learned per-channel scaling generally does not commute with the rotation, since $\operatorname{diag}(\boldsymbol{\gamma})\mathbf{Q}\neq\mathbf{Q}\operatorname{diag}(\boldsymbol{\gamma})$. Consequently, directly applying the matched RoR reparameterization would alter the normalization operation rather than preserve its function. \rev{Protecting these parameters therefore requires a separate mechanism and remains outside the current RoR construction.}

\subsection{Extension to Multimodal Models}

The current evaluation is limited to text-based LLMs. Extending RoR to multimodal models requires validating whether the activation-channel sensitivity estimated from offline calibration remains stable across modalities, since visual and textual inputs may induce different activation distributions. If substantial cross-modal shifts occur, modality-aware calibration or mixing policies may be required. We leave this extension to future work.
\section{Related Work}
\label{sec:related_work}

Bit-Flip Attacks (BFAs) and their defenses have been studied extensively for DNNs and, more recently, Transformer models. We summarize recent BFA attack surfaces, detection- and hardening-based defenses, together with rotation-based quantization methods that are methodologically related to RoR.

\rev{\noindent\textit{1) Recent BFA Attack Surfaces.}
Classical targeted BFAs use model parameters and gradients to identify vulnerable weight bits~\cite{rakin2019bit,yao2020deephammer}. Recent studies have expanded the attack surface beyond conventional weight payloads. Flip-S shows that corrupting quantization scale factors can simultaneously perturb groups of de-quantized weights~\cite{Wang_2025_CVPR}, while Flip-Agent demonstrates targeted BFAs against LLM-based agents, affecting both responses and tool invocations~\cite{wang2026flipagent}. These works highlight distinct BFA surfaces; our study focuses specifically on the Transformer Linear INT8 weight-bit domain defined in Sec.~\ref{subsec:threat_model}.}

\noindent\textit{2) Detection-Based Monitoring.}
Detection-based defenses use runtime monitors, redundancy, or error-correction mechanisms to identify corrupted parameters. Representative methods include NeuroPots~\cite{liu2023neuropots}, ModelShield~\cite{guo2021modelshield}, WeightSentry~\cite{abumandour2025weightsentry}, Aegis~\cite{wang2023aegis}, anomaly-detection schemes~\cite{wen2025anomaly}, and software-based ECC approaches~\cite{liu2024alberta,ahmed2024nn,guan2019place,lee2022value}. RADAR~\cite{li2021radar} instead uses group-level integrity signatures and recovery. Such runtime checking introduces additional operations, while compact signatures may admit checksum-preserving multi-bit collisions under defense-aware attacks. RoR follows a complementary approach by reducing the effect of corrupted Linear weights without relying on runtime integrity detection.

\noindent\textit{3) Parameter-Hardening Methods.}
Parameter-hardening approaches include robust retraining or fine-tuning, such as SAR~\cite{zhou2024sar} and WRecon~\cite{li2020defending}, and encoding-based protection such as DeepNcode~\cite{velvcicky2024deepncode}, CodeNet~\cite{dutta2019codenet}, and RREC~\cite{liu2022generating}. FaR~\cite{nazari2024forget} provides a training-free alternative by rewiring vulnerable Transformer connections to redistribute their functional importance. \rev{We evaluate FaR under the unified LLM deployment protocol in Sec.~\ref{subsec:exp_efficiency}, where implementation and platform differences from its original evaluation are discussed explicitly.}

\rev{\noindent\textit{4) Rotation-Based Quantization Methods.}
Orthogonal rotations have been extensively explored for improving LLM quantization rather than for defending against bit-flip attacks. QuIP~\cite{chee2023quip} applies random orthogonal preprocessing to improve incoherence, QuaRot~\cite{ashkboos2024quarot} uses randomized rotations to suppress activation outliers and improve low-bit quantization, and SpinQuant~\cite{liu2024spinquant} optimizes function-preserving rotations from calibration data to reduce quantization error. These methods are not designed as BFA defenses. Nevertheless, they are methodologically relevant because they also transform the activation/weight representation through orthogonal rotations. RoR differs in both objective and construction: it uses activation-channel sensitivity to build a Householder-motivated hierarchical rotation specifically for robustness against localized weight-bit corruption. We therefore include QuaRot-Had, implemented by following the Hadamard rotation and deployment pattern in the official QuaRot codebase, as a rotation-based reference for separating the effect of generic orthogonal rotation from RoR's BFA-oriented sensitivity-aware design.}
\section{Conclusion}
\label{sec:conclusion}

This paper studies the vulnerability of quantized Large Language Models (LLMs) to bit-flip corruption within Transformer Linear weights. Our analysis shows that severe failures can arise when corrupted weights are coupled to high-sensitivity activation channels, motivating a defense that reduces the concentration of these channel contributions. \rev{We propose Rotated Robustness (RoR), a training-free sensitivity-aware hierarchical rotation motivated by Householder dispersion and realized through structured butterfly mixing. RoR preserves the original Linear mapping in exact arithmetic before re-quantization, while its finite-precision impact is evaluated empirically under INT8 deployment.}

\rev{Across eight evaluated LLMs, RoR achieves the largest or tied-largest observed PPL$>100$ threshold-crossing budget within the PBS100 evaluation horizon under gradient-guided attacks. Under BER-based random corruption, no RoR failures are observed in the evaluated 1,000-trial experiments for the tested models, and the downstream evaluation shows that the resulting robustness also transfers to task-level utility under replayed PBS perturbations. Defense-aware reconstruction further shows that RoR delocalizes specified one-bit catastrophic faults, increasing their reconstruction cost to thousands of deployed INT8 bit changes without implying a global attack lower bound. The unified H200 evaluation additionally shows that this robustness can be obtained with modest persistent storage and runtime overhead.}

Overall, the results indicate that sensitivity-aware redistribution of vulnerable channel contributions is an effective mechanism for improving bit-flip robustness. \rev{Within the evaluated Transformer Linear INT8 weight-bit domain, RoR provides a practical training-free approach that combines targeted robustness, downstream utility preservation, and deployable inference cost.}




\section*{Acknowledgment}
The authors used ChatGPT (OpenAI) to assist with English-language editing and LaTeX formatting.

\bibliographystyle{IEEEtran}
\bibliography{ref}

\begin{IEEEbiography}
[{\includegraphics[width=1in,height=1.25in,clip,keepaspectratio]
{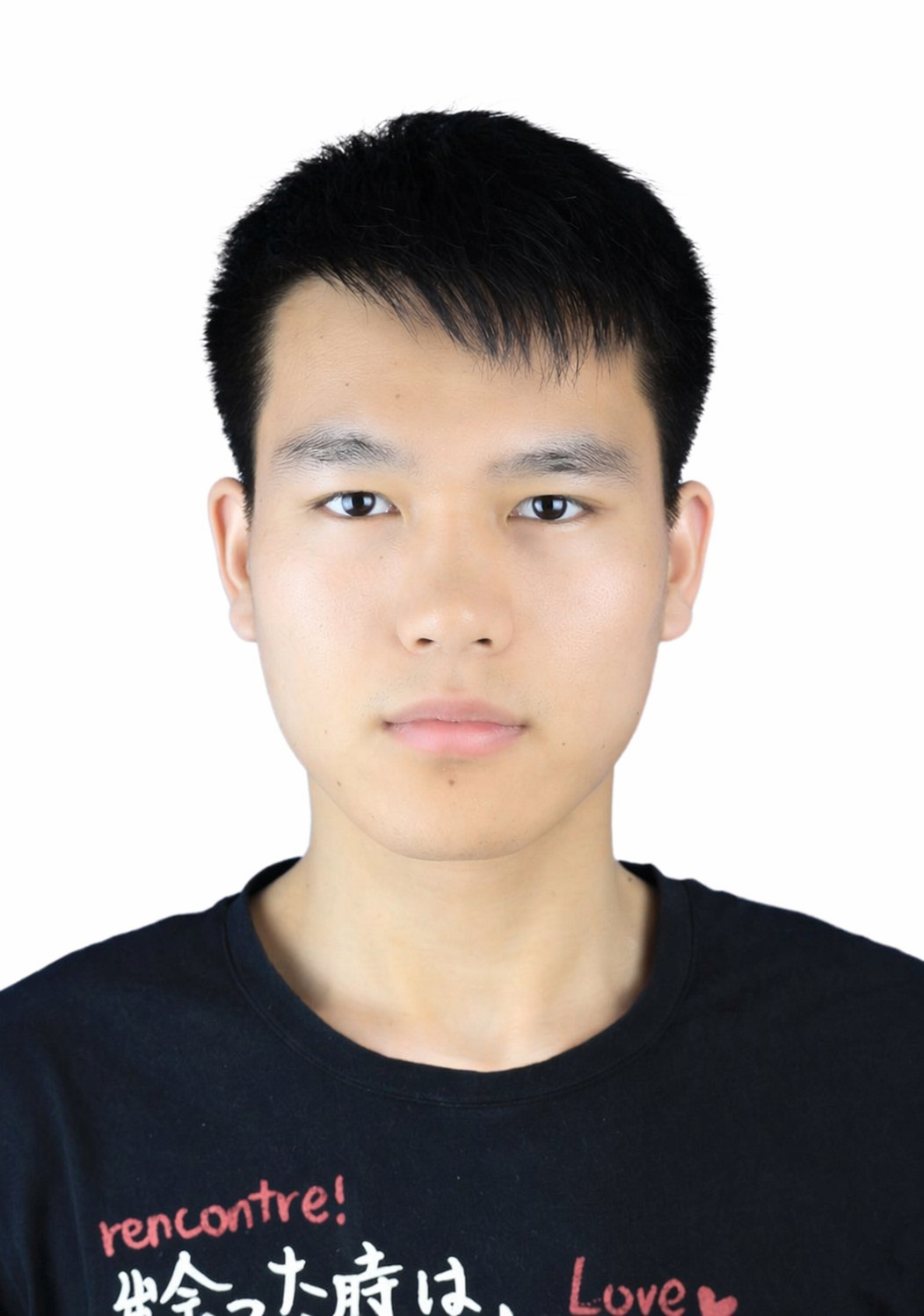}}]{Deng Liu}
(Student Member, IEEE) received the B.S. degree in
microelectronics from the University of Science and Technology of
China (USTC), Hefei, China, in 2021. He is currently pursuing the
Ph.D. degree in microelectronics with the School of Integrated
Circuits, USTC. His research interests include the security and
robustness of large language models, bit-flip attacks and defenses,
inference optimization, and the runtime security of LLM-based agent
systems.
\end{IEEEbiography}

\begin{IEEEbiography}
[{\includegraphics[width=1in,height=1.25in,clip,keepaspectratio]
{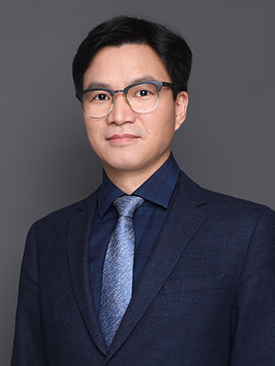}}]{Song Chen}
(Member, IEEE) received the B.S. degree in computer science from Xian Jiaotong University, China, in 2000, and the Ph.D. degree in computer science from Tsinghua University, China, in 2005. He served at the Graduate School of Information, Production and Systems, Waseda University, Japan, as a Research Associate, from August 2005 to March 2009 and as an Assistant Professor, from April 2009 to August 2012. He is currently an Associate Professor with the School of Integrated Circuits, University of Science and Technology of China (USTC). His research interests include several aspects of VLSI design automation, on-chip communication system, in-memory computing, and computer-aided design for emerging technologies. He is a member of ACM and IEICE.
\end{IEEEbiography}

\end{document}